\documentclass[sigconf,nonacm]{acmart}
\AtBeginDocument{%
  }

\setcopyright{acmlicensed}
\copyrightyear{2018}
\acmYear{2018}
\acmDOI{XXXXXXX.XXXXXXX}
\acmConference[Conference acronym 'XX]{Make sure to enter the correct
  conference title from your rights confirmation email}{June 03--05,
  2018}{Woodstock, NY}
\acmISBN{978-1-4503-XXXX-X/2018/06}

\begin{document}

\title{How Do Professional Editors Evaluate the Editing Quality of AI-Generated Cinematic Video Ads?}

\author{Po-Ming Law}
\affiliation{%
  \institution{Adaptive Machines, Inc}
  \city{San Francisco}
  \state{California}
  \country{USA}}
\email{terrance@adaptivemachines.ai}

\author{Weizhi Li}
\affiliation{%
  \institution{Adaptive Machines, Inc}
  \city{San Francisco}
  \state{California}
  \country{USA}}
\email{weizhi@adaptivemachines.ai}

\author{Arpit Narechania}
\affiliation{%
  \institution{The Hong Kong University of Science and Technology}
  \city{Hong Kong}
  \country{China}}
\email{arpit@ust.hk}

\renewcommand{\shortauthors}{Law et al.}

\begin{abstract}
On social media, we often encounter short-form video ads that employ cinematic editing techniques to evoke an emotional response. While AI tools are beginning to generate such cinematic ads automatically, we lack a fine-grained framework for evaluating these ads. In this paper, we first characterize social media video ad formats and identify cinematic ads as a recurring format in our corpus. We then analyze cinematic ads’ duration, shot structure, audio / text elements, and editing techniques to inform a two-stage generation pipeline where an LLM first generates a shot plan and a video generation model renders the video. Using this pipeline, we generated 70 cinematic ads for 35 real brands and recruited professional video editors to critique their editing choices. From their critiques, we derive six dimensions of editing quality: narrative progression, audiovisual coordination and sound design, visual composition and graphics, shot-to-shot continuity, message and brand coherence, and temporal rhythm and pacing. We discuss how these dimensions can guide editing-aware generation, human evaluation, and automated evaluation of AI-generated cinematic ads. \textbf{Supplementary materials:} \url{https://drive.google.com/drive/folders/14FsEjQsnpkwsV4h3HYev_SAFgT5d17v3}
\end{abstract}


\begin{teaserfigure}
 \includegraphics[width=\textwidth]{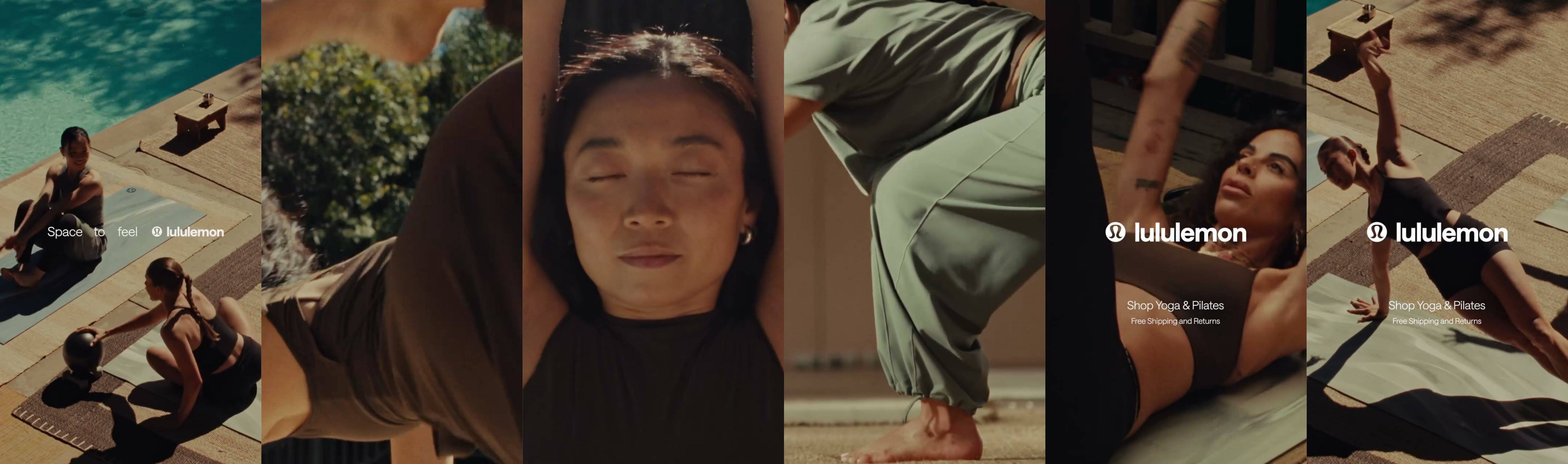}
  \caption{Screenshots of a six-second cinematic video ad from Lululemon's ``Space to feel'' campaign taken one second apart. It shows a montage of women practicing yoga and Pilates outdoors. The first shot is a wide shot of two women. It is followed by a series of close-ups. The color palette feels warm and earthy. The audio features soft music and the sound of a deep breath. This composition aims to evoke a sense of peace and mindfulness.}
  \Description{Six video frames arranged from left to right. The sequence begins with a wide shot of two women exercising outdoors and moves to close-ups of women practicing yoga and Pilates, including a woman's closed eyes and the texture of her clothing. The scenes use warm, golden-hour lighting and earthy colors.}
  \label{fig:lululemon}
\end{teaserfigure}

\received{20 February 2007}
\received[revised]{12 March 2009}
\received[accepted]{5 June 2009}

\maketitle

\section{Introduction}

While browsing social media (e.g., Instagram and Facebook), we routinely encounter short-form video ads. Figure~\ref{fig:lululemon} is a six-second video ad from Lululemon's ``Space to feel'' campaign. It shows a montage of women practicing yoga and Pilates outdoors. It fluidly moves from a wide shot of two women to a series of close-ups (e.g., closed eyes and fabric texture). With the golden-hour natural lighting, the color palette feels warm and earthy. The audio features soft music and the sound of a deep breath. The audiovisual composition evokes a sense of peace and mindfulness.

Ads of this kind employ cinematic editing techniques and often use music and narration to arouse an emotional response. While some other ads may tell a story about a situation and how a product resolves the situation, this class of ads does not have such a traditional narrative arc. In this paper, we use the term cinematic ads to refer to these emotion-arousing ads produced with cinematic editing techniques and lacking a traditional narrative arc.

Advertisers are increasingly using AI to generate these ads~\cite{10.1145/3613905.3636315}. Commercial tools (e.g.,~\cite{creatify, arcads, omneky2026video, tiktok2026symphonyagent, topview2026adgenerator, heygen2026videoagent, amazonads2026creative}) allow users to enter a text prompt with image assets. An AI agent will then produce a complete video ad automatically. These agents can perform sophisticated forms of editing: they can insert captions, connect shots with aesthetically pleasing transitions, and arrange shots into a coherent narrative. Anyone who has watched a stream of AI-generated ads will often recognize something in the editing that feels subtly wrong. However, it is often not easy to articulate what feels off without an appropriate vocabulary.

To accurately evaluate cinematic ads, we require fine-grained dimensions to critique editing quality. Such a fine-grained framework will provide the vocabulary to move past vague impressions (e.g., ``something feels off''). It allows evaluators to identify specific issues (e.g., ``the music changes too frequently,'' ``the motion transitions between the two shots are not smooth,'' and ``the narrative lacks coherence'').

Some frameworks have been proposed for evaluating AI-generated video. Earlier frameworks emphasize low-level properties such as whether objects obey gravity~\cite{pmlr-v267-meng25c} and whether a character's appearance remains consistent across shots~\cite{he-etal-2024-videoscore}. More recently, researchers have begun evaluating the cinematic quality of AI-generated films~\cite{2025arXiv250618899H,filmbench}. Yet, these frameworks do not address the unique editing requirements of AI-generated ads. Ads introduce additional commercial considerations. For example, retaining a shot containing a Nike logo in an AI-generated Adidas ad would violate brand consistency even if the video exhibited high cinematic quality. To derive fine-grained dimensions for evaluating the editing quality of ads, we therefore examined how professional video editors critique AI-generated cinematic ads.

As a first step, we characterized the landscape of social video advertising (Study 1a). We curated 124 video ads from 15 brands across 5 product categories. Grounded in our analysis, we identified eight recurring ad types: cinematic ads, user-generated content (UGC), product demos, slideshows, problem-solving narratives, visual hooks, monologues, and interviews. We found that cinematic ads emerged as one of the most prevalent formats in our corpus.

From Study 1a, we collected 36 cinematic ads. In Study 1b, we additionally collected 63 cinematic ads and analyzed the duration, shot count, audio and text elements, and editing techniques used in these 99 cinematic ads. Our analysis suggests that cinematic ads on social media constitute a form distinct from television ads and movies. Compared with television ads, cinematic ads tend to be shorter. Compared with movies, cinematic ads have a shorter shot duration and employ editing techniques that are less frequently used in movies.

Studies 1a and 1b defined what cinematic ads are. With this definition, we asked: How do professional video editors evaluate AI-generated cinematic ads? To investigate this question, we developed a two-step pipeline for generating these ads. In the first step, users provide a brand description and assets. An LLM then creates a shot plan. The shot plan has a sequence of shots with descriptions. It depicts the editing within each shot and the transitions between shots. A video generation model then uses the shot plan to produce a complete cinematic ad. We created 70 cinematic ads for 35 brands using this pipeline.

We recruited six professional video editors to evaluate these 70 cinematic ads. Through an online interface, we asked editors to provide open-ended critiques of what worked, what failed or was missing, and how they would edit each ad differently. After the online critique, we conducted a one-hour interview with each editor to disambiguate the written responses and discuss editing considerations. We derived six dimensions along which professionals evaluate the editing quality of AI-generated cinematic ads: (1) narrative progression, (2) audiovisual coordination and sound design, (3) visual composition and graphics, (4) shot-to-shot continuity, (5) message and brand coherence, and (6) temporal rhythm and pacing.

Our framework is crucial for improving the editing quality of AI-generated cinematic ads. First, it guides future research directions. For example, Study 2 identifies the editing shortcomings of video generation systems for future improvement. Second, it provides a foundation for developing more structured forms of human evaluation. For example, with the six dimensions we identified, researchers can develop rubrics for evaluating the editing quality of cinematic ads. Human raters can score an AI-generated cinematic ad using the rubrics. Third, the dimensions could inform future automated evaluators or autoraters designed to assess editing quality in ways that better reflect professional judgment.

The contributions of this paper are two-fold:
\begin{itemize}
\item  A characterization of short-form cinematic video ads on social media.
\item  A professional-editor-derived framework consisting of six editing-quality dimensions for critiquing AI-generated cinematic ads.
\end{itemize}

\section{Related Work}

In Study 1a, we drew on existing research regarding video ad formats to categorize video ads on social media. In Study 1b, we analyzed video editing techniques within cinematic ads. We gathered these techniques from the film editing literature. Study 2 derived dimensions for evaluating AI-generated cinematic ads from video editors' critiques of such ads. This study is situated within research on video editing practices and has implications for developing video generation benchmarks.

Accordingly, we review related work on video ad formats, video editing techniques, video editing practices, and video generation benchmarks.

\subsection{Video Ad Formats}

Marketing research has a long tradition of studying the impact of different narrative structures on the comprehensibility, likability, and persuasiveness of television ads. For example, van Enschot and Hoeken~\cite{542be46c-a572-38d4-af90-9f868d476a91} showed that whether a visual metaphor is verbally or visually explained affects how comprehensible a television ad is.

Within this tradition, researchers often characterize television ad formats before investigating the differences among formats. Stern~\cite{d5b5dba5-a060-358d-bcca-a85d7586b5b3} identified two types of television ads: classical dramas and vignette dramas. Classical dramas follow the beginning, turning point, and resolution structure described by Freytag's Pyramid~\cite{freytag1894technique}. A television ad with this form may start with a problem and conclude with the product solving that problem. Vignette dramas, on the other hand, present a series of loosely associated stories.

Similarly, Martínez et al.~\cite{5faf31ab29995257011955d8} have characterized the narrative structure of television ads. They analyzed 200 television ads using Ryan’s framework of narrativity~\cite{7674a198-fb17-3e03-98bd-7c7640b83522}. They observed that these ads often have a problem-solving narrative structure. Such ads project a problem state that is ultimately resolved by the advertised products. 

In Study 1a, two of the formats we observe on social media align with classical and vignette dramas identified by Stern~\cite{d5b5dba5-a060-358d-bcca-a85d7586b5b3}: our problem-solving narrative format is a type of classical drama as it has a problem-product-resolution arc; our cinematic format shares the structure of the vignette drama since it often presents a montage of related imagery without a problem-solving narrative structure.

Our work complements this marketing literature in two ways. In Study 1a, we classified social media video ads as opposed to television ads. We identified eight recurring ad formats. Some, such as visual hooks, have no televised counterpart. In Study 1b, we analyzed the duration, shot count, audio and text elements, and editing techniques in cinematic ads. We observed that cinematic ads on social media are a format distinct from television ads and movies. For example, the duration of cinematic ads tends to be shorter than that of television ads.

\subsection{Video Editing Techniques}

Many online resources (e.g.,~\cite{studiobinder2021transitions}) offer tutorials about cinematic editing techniques. Seminal works in film editing theory explore the art of editing and provide rich vocabularies for editing devices. For example, Murch's In the Blink of an Eye proposes the ``Rule of Six,'' which is a set of six criteria that make a cut feel seamless~\cite{murch2001blink}. Arijon's Grammar of the Film Language catalogs editing conventions such as eyeline matches~\cite{arijon1976grammar}. Burch's Theory of Film Practice systematizes transition types by their spatial and temporal articulations~\cite{burch1981theory}. It discusses techniques like the straight match cut (e.g., cutting from someone turning a doorknob to a reverse-angle shot of them walking through the door) and temporal ellipses (e.g., a character starts climbing stairs in one shot and is already on the fifth floor in the next).

Notably, there is no consensus on a complete inventory of video editing techniques. As Murch notes, video editing is a creative practice whose vocabulary of devices continuously expands and evolves~\cite{murch2001blink}. Similarly, while defining a taxonomy of cut types for recognition, Pardo et al.~\cite{10.1007/978-3-031-20071-7_39} acknowledge that their list of cut types is partial. Hence, we do not claim to exhaustively enumerate the editing techniques of cinematic ads. Instead, in Study 1b, we identified a set of editing techniques based on existing literature and our observations after reviewing social media cinematic ads. Using this set, we analyzed the presence of editing techniques across our sample of cinematic ads. 

\subsection{Video Editing Practices}

Video production commonly involves repeated cycles of editing a video, providing feedback, and revising the video~\cite{vidcrit}. In Study 2, we specifically focus on what feedback professional editors provide concerning the editing quality of AI-generated ads. Hence, our study is situated within the broader context of research on video editing practices.

HCI research has studied video editing workflows and developed interfaces for specific editing activities. Such activities include organizing clips into chunks~\cite{chunkyedit}, matching sound effects to video~\cite{soundify}, and transforming long-form videos into short-form videos~\cite{lotus}. Some researchers have also developed interfaces to enhance the ability of editors to express their editing intents (e.g.,~\cite{expressiveedit, avscript}). For example, ExpressEdit~\cite{expressiveedit} combines natural language and sketching for specifying edits.

More closely related to our study is research that has studied how reviewers communicate feedback on edited videos and how reviewers address the feedback (e.g.,~\cite{collaborative-design, collaborative-unpack, vidcrit}). For example, Pavel et al.~\cite{vidcrit} interviewed video authors to learn about how they give and receive feedback. They then developed an interface that supports collaboratively recording feedback and viewing feedback asynchronously.

Researchers have also examined how video creators work with AI (e.g.,~\cite{creator-ai, videodiff}). For example, Huh et al.~\cite{videodiff} conducted a formative study with video editors, during which the researchers observed how editors compared variations of the same video. They then developed VideoDiff, in which AI generates multiple variations and users review the AI recommendations.

Our work complements these studies by investigating what feedback video editors give regarding the editing quality of AI-generated ads. From the reviews, we derive dimensions for evaluating the editing quality of these ads. The resulting dimensions provide a vocabulary that future human--AI co-creation systems could use to organize feedback and support targeted revisions.

\subsection{Video Generation Benchmarks}

Benchmarks for assessing video quality predate generative AI. For example, Tu et al.~\cite{Tu_2021} developed an evaluator for assessing the quality of UGC created by human creators. They demonstrated that their evaluator produced video quality scores that correlate well with human judgment.

With the advent of video generation models, newer benchmarks have focused on evaluating the quality of AI-generated videos (e.g.,~\cite{Liu_2023, FAN2023100152, Wang_2025, pmlr-v267-meng25c, Sun_2025, evalcrafter}). These benchmarks aim to compare the quality of videos generated by multiple video generation models. During an evaluation, a prompt set is used to generate videos from each model. For each generated video, we calculate scores along specific dimensions. We then compare models by comparing the aggregated scores of their generated videos.

Many benchmarks assess low-level properties of AI-generated videos such as realism and plausibility. VBench defines metrics for video quality (e.g., whether a character's appearance remains consistent throughout the video) and video-condition consistency (e.g., whether the video correctly generates objects described in the text prompt)~\cite{Huang_2024}. VBench-2.0 proposes five dimensions for evaluating intrinsic faithfulness: human fidelity, creativity, controllability, commonsense, and physics~\cite{2025arXiv250321755Z}.

More recently, some researchers have developed benchmarks for assessing the cinematic quality of AI-generated films (e.g., FilmEval~\cite{2025arXiv250618899H} and FilmBench~\cite{filmbench}). For example, Wang et al. collaborated with professional directors and film-school faculty to develop FilmBench's three-level evaluation taxonomy. It comprises three axes: instruction following, temporal continuity, and aesthetic quality. These axes are further decomposed into 12 components and 35 sub-metrics.

While FilmEval~\cite{2025arXiv250618899H} and FilmBench~\cite{filmbench} also evaluate the use of cinematic techniques, their focus is strictly on films, not ads. From Study 1b, we learned that cinematic ads use editing techniques that might be used less frequently in films. From Study 2, we derived message and brand coherence as an evaluation dimension for cinematic ads. This dimension concerns whether the shots convey a clear central message and accurately portray the brand image. Message and brand coherence is generally inapplicable to traditional films. This highlights why cinematic ads warrant a distinct set of dimensions for evaluating editing quality.

\section{Study 1a: Characterizing Social Media Ad Formats}

What are the formats of video ads brands run on social media? To find out, we collected 124 social media ads from the Meta and TikTok ad libraries and conducted a qualitative analysis of this corpus.

\subsection{Methodology}

\subsubsection{Data Collection}

We collected video ads from five consumer-facing product categories: apparel and footwear, beauty and personal care, consumer technology, direct-to-consumer e-commerce, and food and beverage. We selected these categories to capture visual variations and diversity. For instance, apparel and footwear ads often foreground fabric and bodily movement; beauty and personal care ads may rely on face close-ups and hands-on product application; food and beverage ads often emphasize sensory appeal and consumption moments. For each category, we selected three brands. Hence, we collected ads from 15 brands (we provide the full list in the supplementary materials). The data collection took place between May and July 2026.

We retrieved these video ads from the Meta~\cite{meta_ad_library} and TikTok~\cite{tiktok_ad_library} ad libraries, which house ads posted across Facebook, Instagram, and TikTok. For each brand, we entered the brand name as a keyword and sorted the ads in descending order of impressions (i.e., how many times an ad is viewed). From each ad library, we selected at most five top-ranked ads. We focused on the ads in English and filtered out the ones that were not. We further removed duplicates. This is because the same video ad can appear multiple times within an ad library and across ad libraries. This procedure yielded 124 unique video ads.

\subsubsection{Analysis}

With the 124 ads, an author open-coded the full set to develop a preliminary codebook. This codebook contains the definition, inclusion criteria, and exclusion criteria for each ad format and is provided in the supplementary materials. Next, two coders independently coded 25\% of the data randomly selected from the full set. We discussed disagreements, resolved inconsistencies, and refined the codebook. We then drew 25\% of the data from the uncoded sample for a subsequent round of independent coding. We repeated this process until inter-rater reliability reached a Cohen's $\kappa$~\cite{mchugh2012interrater} above 0.7. With the final codebook, one coder completed the remaining sample.

\subsection{Social Media Ad Formats}

Eight ad formats emerged from our analysis. In the following, we describe them in descending order of frequency in our corpus.

\begin{figure*}
  \includegraphics[width=\textwidth]{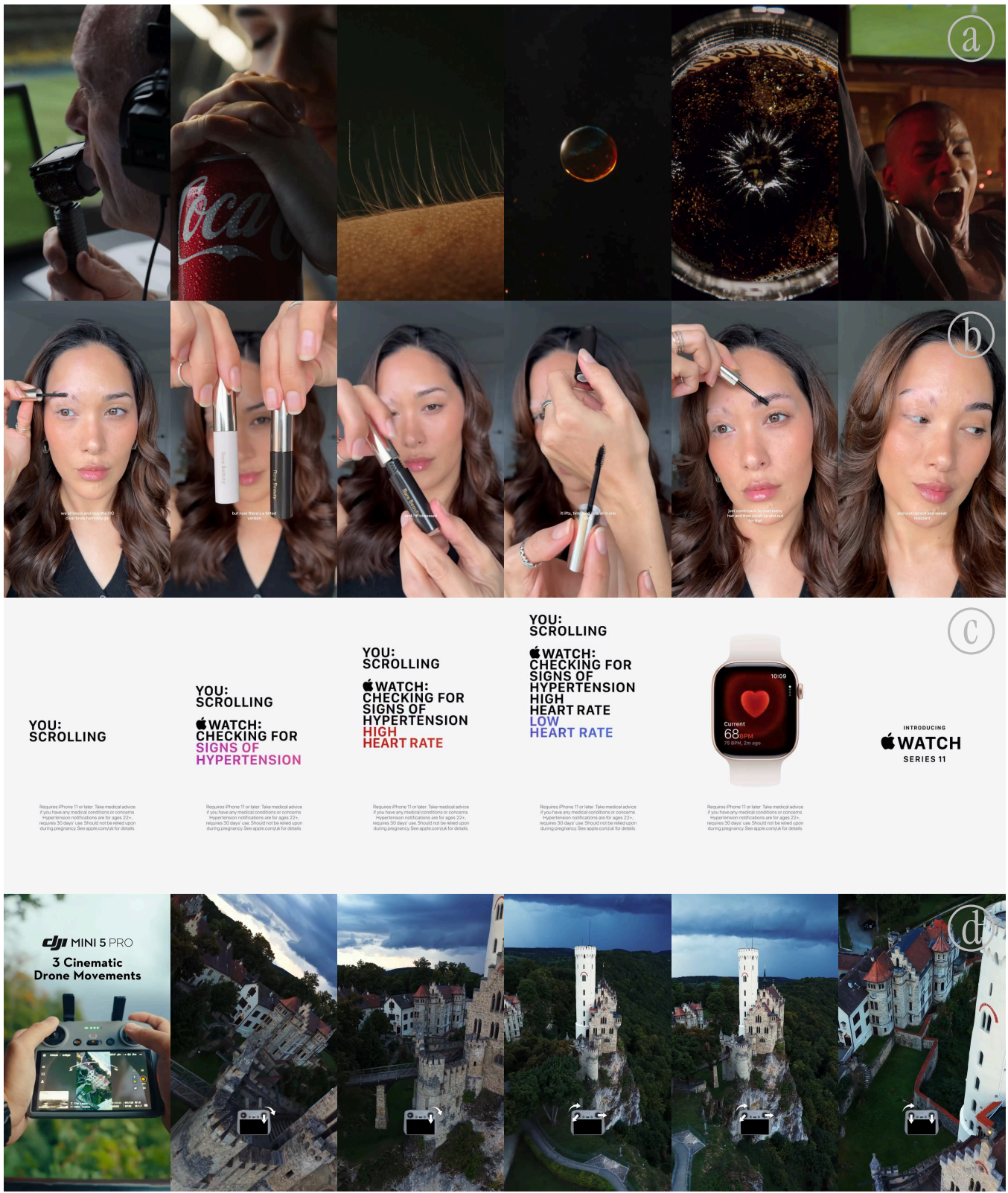}
  \caption{Selected key frames in a Coca-Cola cinematic ad (a), a Rare Beauty UGC-style ad (b), an Apple slideshow ad (c), a DJI product demo ad (d).}
  \Description{Four labeled groups of video frames illustrate different ad formats. Group a shows a Coca-Cola cinematic ad progressing from a sports commentator and nervous football fans to a match cut between moving hair and a rising soda droplet, followed by celebrating fans. Group b shows a woman applying Rare Beauty brow gel in a vertically framed home video. Group c shows an Apple Watch slideshow composed of text cards and a product image. Group d shows aerial DJI drone footage paired with controller graphics illustrating joystick movements.}
  \label{fig:ad_set_1}
\end{figure*}

\subsubsection{Cinematic (36 ads, 29\% of ads in our corpus)}

Cinematic ads employ cinematic editing techniques (e.g., match cuts). They arouse an emotional response through imagery, music, and narration. Traditional narrative arcs in ads present a situation that is resolved by the product. Cinematic ads lack an explicit problem-and-resolution arc.

Figure~\ref{fig:ad_set_1}a shows a Coca-Cola ad from the ``Drink in the FIFA World Cup 26'' campaign. This ad does not have a problem-and-resolution arc. The first shot shows a close-up of a sports commentator in a stadium (Figure~\ref{fig:ad_set_1}a frame 1). The commentator's voice and the ambient sounds of the match persist throughout the video. A montage of nervous football fans follows the first shot (Figure~\ref{fig:ad_set_1}a frame 2).

When the commentator mentions an impending \textit{``free kick,''} the film utilizes a motion match cut. The camera captures hair moving upward in one shot (Figure~\ref{fig:ad_set_1}a frame 3), which then seamlessly transitions into a macro shot of a golden soda droplet moving upward (Figure~\ref{fig:ad_set_1}a frame 4).

As the soda droplet falls into the larger pool of liquid (Figure~\ref{fig:ad_set_1}a frame 5), the ad signals the scoring of a goal. This is followed by a burst of excitement associated with scoring a goal. We see a rapid montage of football fans reacting in jubilant cheers (Figure~\ref{fig:ad_set_1}a frame 6).

\subsubsection{UGC (30 ads, 24.2\%)}

UGC-style ads have raw footage with casual editing. They showcase some products or share a personal experience using a first-person angle. These ads can adopt different presentation formats: a content creator speaking directly to the camera, an off-screen creator narrating the video throughout, or a silent visual demonstration that relies entirely on captions written from the user's point of view.

Figure~\ref{fig:ad_set_1}b shows a Rare Beauty ad. This ad is filmed in a home environment with a vertical selfie framing. The content creator demonstrates the application of a tinted brow gel. Although she does not speak directly to the camera, she provides a first-person voiceover narration—expressing her personal obsession with the product while guiding the audience through her step-by-step application process.

\subsubsection{Product Demo (28 ads, 22.6\%)}

In contrast to UGC, product demo ads highlight a product's features or functionality from an objective, third-person perspective. For technical devices or gadgets, these ads often adopt an instructional approach to explain how the product operates. Conversely, when showcasing apparel, a product demo might feature a model wearing various outfits within a single setting.

Figure~\ref{fig:ad_set_1}d is a DJI ad that demonstrates three cinematic drone movements of the DJI Mini 5 Pro. It is structured into three parts, each demonstrating a drone movement (Figure~\ref{fig:ad_set_1}d frames 2 - 3 are part 1; frames 4 - 5 are part 2; frame 6 is part 3). Each part has the drone’s aerial footage with a controller graphic at the bottom of the frame showing the joystick inputs that produce that camera movement. This ad feels instructional by showing the product’s capabilities while teaching the viewer how to achieve them.

\subsubsection{Slideshow (16 ads, 12.9\%)}

Slideshow ads convey their message through a sequence of text-heavy visual cards. There are often animated transitions between cards.

Figure~\ref{fig:ad_set_1}c shows an Apple Watch ad. The ad opens with a text card that says, \textit{``YOU: SCROLLING''} (Figure~\ref{fig:ad_set_1}c frame 1). Utilizing animated transitions, the next card adds a message: \textit{``APPLE WATCH: CHECKING FOR SIGNS OF HYPERTENSION''} (Figure~\ref{fig:ad_set_1}c frame 2). At the end, we see a card with an Apple Watch (Figure~\ref{fig:ad_set_1}c frame 5) and a static concluding card that reads, \textit{``INTRODUCING APPLE WATCH SERIES 11''} (Figure~\ref{fig:ad_set_1}c frame 6).

\begin{figure*}
  \includegraphics[width=\textwidth]{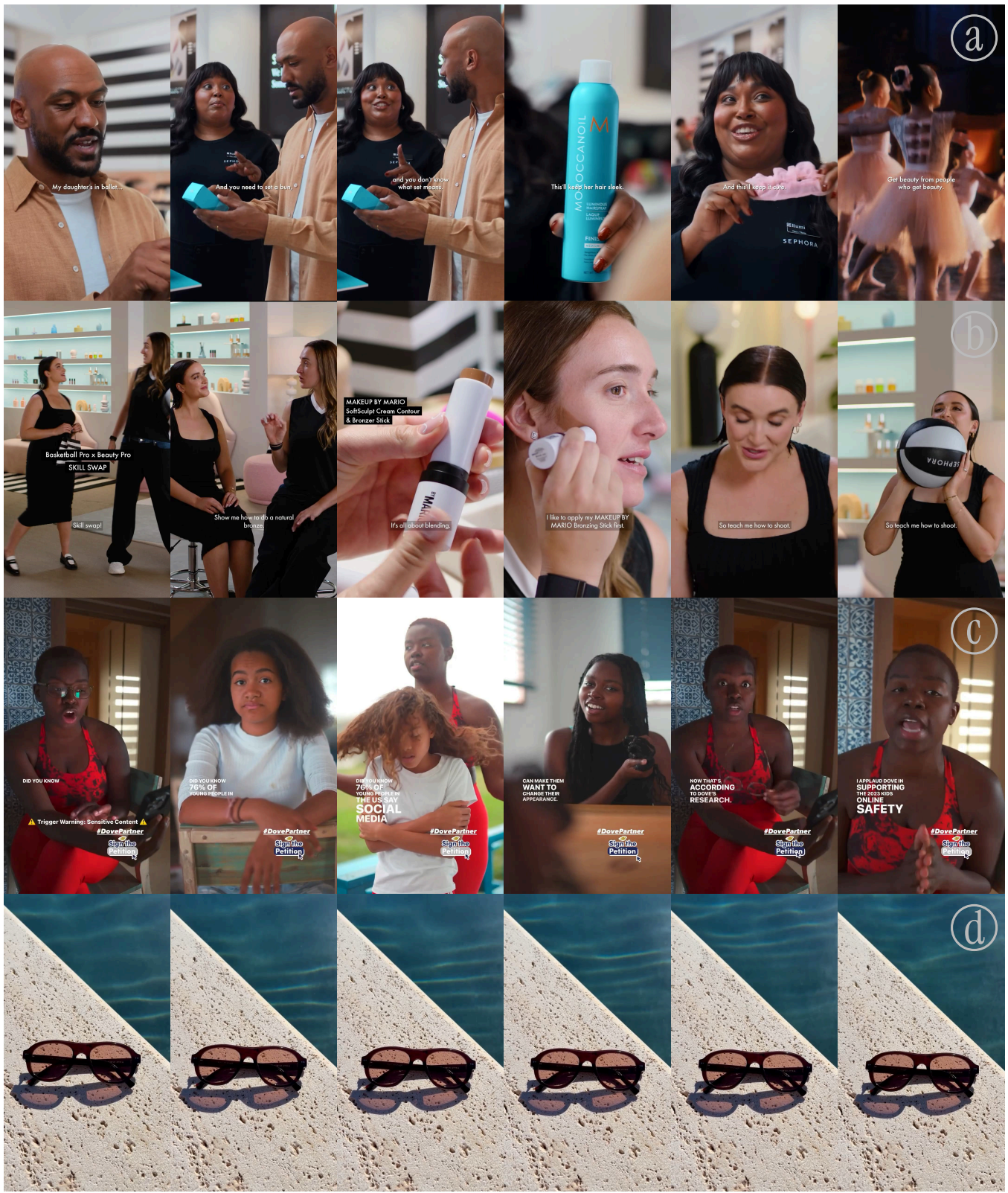}
  \caption{Selected key frames in a Sephora problem-solving ad (a), a Sephora interview ad (b), a Dove monologue ad (c), a Warby Parker visual hook ad (d).}
  \Description{Four labeled groups of video frames illustrate different ad formats. Group a shows a father seeking advice in a Sephora store, receiving hair products, and later watching his daughter perform ballet with her hair in a bun. Group b shows a basketball player and a Sephora beauty advisor teaching each other makeup and basketball skills. Group c alternates between a woman speaking to the camera and cinematic parent-child footage. Group d shows a single static shot of sunglasses resting on the stone edge of a swimming pool.}
  \label{fig:ad_set_2}
\end{figure*}

\subsubsection{Problem-Solving Narrative (5 ads, 4\%)}

These ads follow a traditional storytelling arc: a problem, frustration, or situation is established, and the product or service resolves it. 

Figure~\ref{fig:ad_set_2}a illustrates a Sephora ad. In the ad, a father tells an advisor in a Sephora store that his daughter's ballet class requires \textit{``setting a bun''} and he doesn't know what that means (problem) (Figure~\ref{fig:ad_set_2}a frames 1 - 3). The advisor recommends a hairspray and a scrunchie (intervention) (Figure~\ref{fig:ad_set_2}a frames 4 - 5). The ad cuts to the finished bun and the daughter performing on stage (resolution) (Figure~\ref{fig:ad_set_2}a frame 6). The story suggests that Sephora’s in-store experts can empower consumers to navigate beauty-related challenges.

\subsubsection{Visual Hook (4 ads, 3\%)}

On social media, a video ad is often accompanied by some text message. Visual hooks are ads with a single shot and no cuts, and they tend to be short (e.g., around five seconds). They are designed to be read together with the accompanying text of a social media post. When viewed in isolation, the intended message could be hard to decipher.

For example, Figure~\ref{fig:ad_set_2}d is a Warby Parker ad. It consists of a single shot of sunglasses resting on a stone pool ledge. Based on the video alone, a viewer would struggle to even identify the brand, as the logo is barely legible on the product. However, the ad is contextualized by its accompanying social media caption: \textit{``Our latest wave of summer styles includes new, sun-ready shapes and a refreshed palette.''} By reading this supplementary text, the viewer can understand that the ad serves as a product announcement.

\subsubsection{Monologue (3 ads, 2.4\%)}

Different from UGC, which consists of raw, casually edited footage, monologue ads center a single person speaking to the camera with higher production and editing quality than self-shot videos.

Figure~\ref{fig:ad_set_2}c shows a Dove advocacy ad in which a creator discusses social media's effects on young people's self-esteem. The creator speaks directly to the camera. Her monologue is intercut with B-roll of parent–child imagery (Figure~\ref{fig:ad_set_2}c frames 2 - 4). The slow motion and shallow depth of field lend the B-roll a cinematic quality.

\subsubsection{Interview (2 ads, 1.6\%)}

Interview ads involve question answering between two or more people. The interviewer need not be visible.

Figure~\ref{fig:ad_set_2}b shows a Sephora ``skill swap'' ad pairing a professional basketball player with a beauty advisor. The basketball player first asks, \textit{``Show me how to do natural bronze,''} (Figure~\ref{fig:ad_set_2}b frame 2) and the beauty advisor demonstrates how to do it (Figure~\ref{fig:ad_set_2}b frames 3 - 4). The beauty advisor then requests, \textit{``So teach me how to shoot''} (Figure~\ref{fig:ad_set_2}b frame 5) and the basketball player does so (Figure~\ref{fig:ad_set_2}b frame 6).

\section{Study 1b: Characterizing Cinematic Ads}

Study 1a established cinematic ads as a common format on social media. In Study 1b, we further analyzed the duration, shot count, audio and text elements, and editing techniques in cinematic ads. 

Study 1b also informed the design of our pipeline for producing AI-generated cinematic ads (Section~\ref{sec:generate}). This pipeline uses an LLM to generate a textual shot plan, including shot descriptions, within-shot editing instructions, and transitions between shots. To ground the shot plans in actual advertising practice, we seeded the LLM system prompt with editing techniques observed in Study 1b.

\subsection{Methodology}

\subsubsection{Data Collection}

To obtain a larger set of cinematic ads, we augmented the Study 1a corpus with five additional product categories: athletic and activewear, automotive, beauty and fragrance, beverage and spirits, and fashion and luxury. We selected these categories because brands in these categories appeared to commonly produce cinematic ads. For example, automotive ads often emphasize fluid motion and immersive experience; beverage and spirits ads often use sensory and social imagery; fashion and luxury ads commonly foreground artful composition.

For each category, we selected four brands, resulting in a total of 20 brands. We provide the full list of brands in the supplementary materials. We then gathered video ads for these brands from the Meta and TikTok ad libraries using the same procedure as in Study 1a. This process yielded 175 additional video ads. One coder then applied the Study 1a codebook to the new corpus and identified 63 new cinematic ads. Ambiguous cases were discussed with a second coder. Combined with the 36 cinematic ads identified in Study 1a, the final Study 1b corpus comprised 99 cinematic ads.

\subsubsection{Analysis}

We analyzed four aspects of cinematic ads: duration, shot count, audio and text elements, and editing techniques. We extracted the duration directly from each video’s metadata. We coded shot count manually. Audio and text elements (music, captions, narration, dialogue, and monologue) were binary items (e.g., for music, we coded the presence of it in the ad). Pardo et al.~\cite{10.1007/978-3-031-20071-7_39} suggested that there is no consensus on a complete set of editing techniques. We compiled an initial set of techniques from film editing literature (e.g.,~\cite{murch2001blink, arijon1976grammar, burch1981theory}), online editing resources (e.g.,~\cite{studiobinder2021transitions}), and our own observations while reviewing the cinematic ads. Each editing technique was also coded as a binary variable indicating whether the technique appeared in the ad.

We developed an initial codebook that defines 22 items: a shot, the five audio and text elements, and 16 editing techniques. Using this initial codebook, we coded the cinematic ads following a process similar to Study 1a. For each item, two coders independently coded 25\% of the data randomly selected from the corpus. Since shot count is a continuous variable, we calculated the intraclass correlation coefficient to measure the interrater reliability~\cite{koo2016guideline}. For the 21 binary items, we computed Cohen's Kappa. The coders discussed disagreements and refined the definitions in the codebook. We drew additional 25\% subsets from the uncoded data until reliability was acceptable. For low-frequency items, where kappa can be unstable, disagreements were resolved through discussion. With the final codebook, one author coded the remaining data.

\subsection{Findings}

Here, we summarize our findings from qualitative coding. We provide the full coding results in the supplementary materials.

\subsubsection{Duration, Shot Count, and Shot Duration}

Cinematic ads in our corpus have an average total duration of 17.9 seconds, an average of 15.2 shots, and an average shot duration of 1.2 seconds.

These numbers indicate that cinematic ads on social media tend to have a faster pace than TV ads and films: Television ads run approximately 30 seconds~\cite{pieters1997consumer} while our cinematic ads last 17.9 seconds on average; the average shot duration of modern films is approximately four seconds~\cite{Cutting_2016} while the average shot duration of our cinematic ads is approximately 1.2 seconds.

\subsubsection{Editing Techniques}

The most common techniques include cut on action (present in 52.2\% of our corpus), montage (37.4\%), cut to beat (22.2\%), movement match cut (18.2\%), and overlay-imagery alignment (16.2\%).

We observed that cinematic ads may employ editing techniques that are less commonly used in films. For example, 44.4\% of our corpus has text overlay. Among these ads, 36.4\% employ overlay-imagery alignment: the text overlay aligns with and describes the concurrently shown imagery.

\begin{figure*}
  \includegraphics[width=\textwidth]{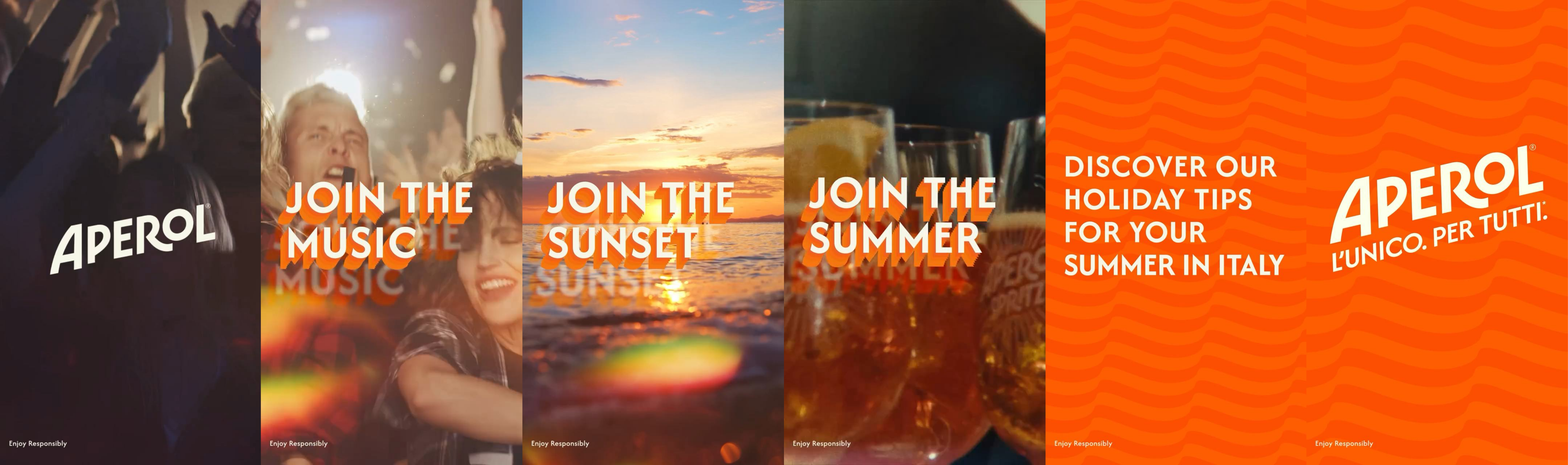}
  \caption{Selected key frames in an Aperol ad that employs overlay-imagery alignment. The ad opens with a shot of a crowd cheering at a music festival with a \textit{``JOIN THE MUSIC''} text overlay (frame 2). It then cuts to a golden-hour sunset over rippling ocean water, pairing the visuals with the text \textit{``JOIN THE SUNSET''} (frame 3). Next, it shows a close-up of Aperol Spritz glasses clinking in a toast, arousing the desire for summer socializing with an overlay reading \textit{``JOIN THE SUMMER''} (frame 4).}
  \Description{A sequence of video frames pairs summer imagery with corresponding text overlays. A cheering festival crowd appears with the words ``JOIN THE MUSIC,'' a sunset over the ocean appears with ``JOIN THE SUNSET,'' and several Aperol Spritz glasses clinking together appear with ``JOIN THE SUMMER.''}
  \label{fig:aperol}
\end{figure*}

For example, Figure~\ref{fig:aperol} presents an Aperol ad that employs overlay-imagery alignment. The ad opens with a shot of a crowd cheering at a music festival (Figure~\ref{fig:aperol} frame 2). The text overlay reads \textit{``JOIN THE MUSIC''} to align with the imagery. It then cuts to a golden-hour sunset over rippling ocean water, pairing the imagery with the text \textit{``JOIN THE SUNSET''} (Figure~\ref{fig:aperol} frame 3). Next, it shows a close-up of multiple Aperol Spritz glasses clinking together in a toast (Figure~\ref{fig:aperol} frame 4). This shot arouses the desire for socializing with friends during the summer and shows an overlay that reads \textit{``JOIN THE SUMMER.''} Throughout, the ad plays upbeat audio and a montage of summer-related shots. The audiovisual experience establishes a vibrant summer vibe.

On the other hand, films do not typically rely on text overlays to deliver the narrative, instead utilizing character action and dialogue to drive the story~\cite{bordwell2016film}. Overlay-imagery alignment could therefore be less common in films.

\subsubsection{Usage of Montage}

We adopted a conservative definition of montage so we could apply the code reliably during the analysis: three or more consecutive shots with different people, objects, and / or scenes that share the same underlying concept. Under this definition, over one-third (37.4\%) of the cinematic ads in our corpus contain a montage. The Lululemon ad in Figure~\ref{fig:lululemon}, the Coca-Cola ad in Figure~\ref{fig:ad_set_1}a, and the Aperol ad in Figure~\ref{fig:lululemon} are examples. Specifically, the Aperol ad strings together shots with different settings (the crowd, the sunset, and the glasses) that collectively evoke a cohesive summer atmosphere.

\section{Generating Cinematic Ads}
\label{sec:generate}

Studies 1a and 1b defined what cinematic ads are. Study 2 elicited professional critiques of AI-generated cinematic ads. We aimed to focus on a fully automated generation paradigm: the user describes their brand, and the system makes the creative decisions needed to produce a complete video. As video generation models become more powerful, this automated paradigm has become increasingly common in commercial ad generators (e.g.,~\cite{creatify, arcads}). To conduct Study 2, we needed a pipeline for generating cinematic ads.

\begin{figure}
  \includegraphics[width=\columnwidth]{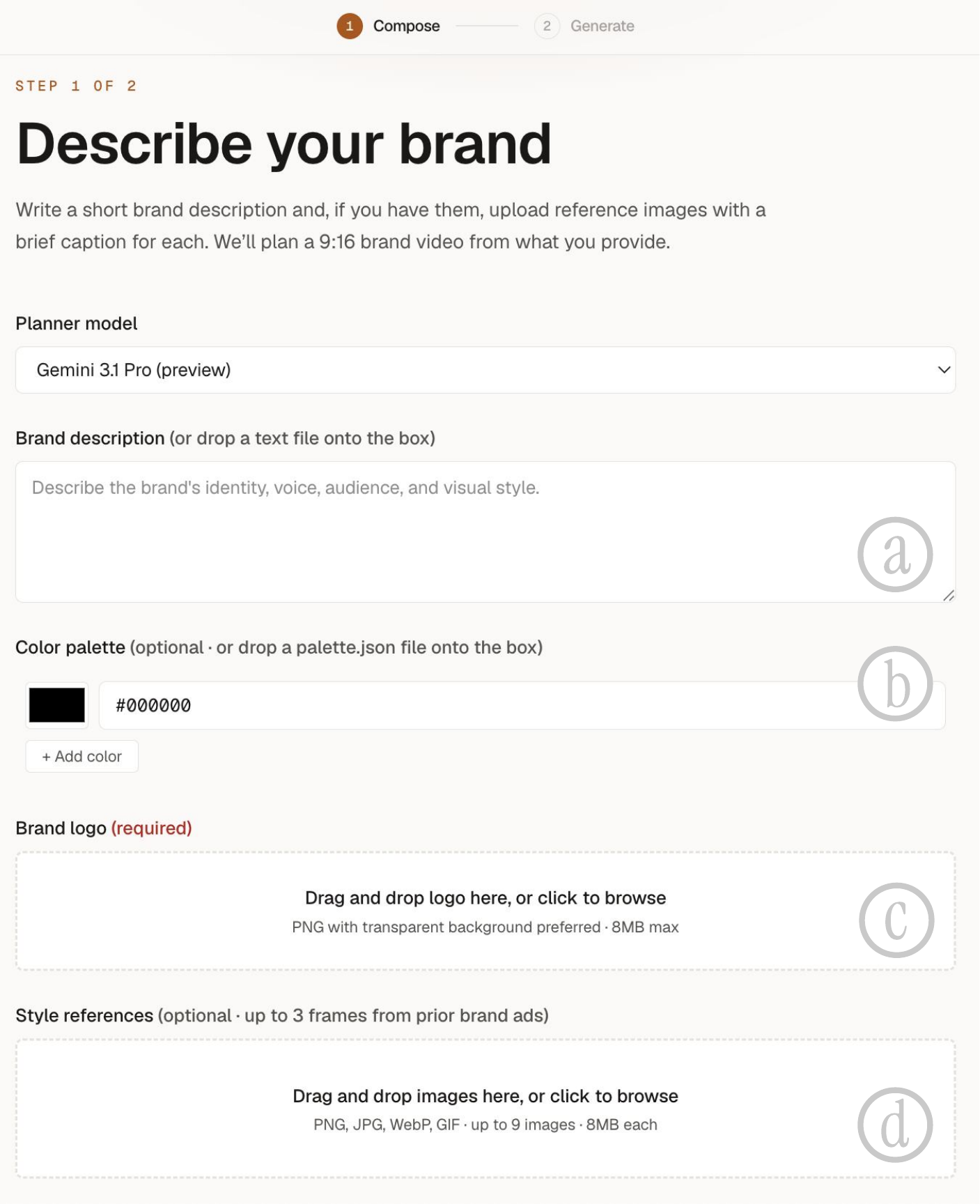}
  \caption{The web interface for generating cinematic ads. It takes brand description (a), color palette (b), logo (c), and style-reference images (d) as inputs. With the brand inputs, an LLM planner produces a shot plan and a video generation model renders the video based on the shot plan.}
  \Description{Screenshot of the cinematic-ad generation interface. The input area contains fields for a brand description, color palette, brand logo, and style-reference images. The interface presents these inputs to an LLM planner that creates a shot plan, which is then passed to a video generation model to render the ad.}
  \label{fig:interface}
\end{figure}

We implemented a web interface (Figure~\ref{fig:interface}) that takes brand description, color palette, logo, and style-reference images as inputs. It then generates ads in two stages. First, an LLM planner produces a shot plan with descriptions about the shots and editing techniques used within and between shots. Second, a video-generation model generates a cinematic ad using the plan.

This two-step workflow mirrors professional video production. During such production, directors and editors often prepare shot lists and storyboards and then produce the footage~\cite{katz1991film}. The two-step workflow is also common in recent AI video generation systems (e.g.,~\cite{2025arXiv251015831L, 10.1007/978-3-031-73027-6_27, videodirector}). Such systems use an LLM to generate a structured shot plan before synthesizing video.

Here, we describe how the LLM generates a shot plan and how the video generation model generates a video from the plan.

\subsection{Generating Shot Plan}

\subsubsection{Brand Inputs}

To generate a shot plan, users provide some brand inputs. Karnatak et al.~\cite{2025arXiv250414320K} interviewed business owners and learned that it is challenging to express brand intuition through textual prompts alone. They proposed a content generation pipeline that takes brand descriptions (e.g., value and audience), color palette, and brand assets (e.g., logo and visual inspirations) for fine-grained control. 

Following their approach, our web interface (Figure~\ref{fig:interface}) allows users to provide brand description, color palette, logo, and style-reference images. The style references could be frames extracted from the brand’s prior advertisements. They aim to communicate the brand’s visual qualities such as color treatment and overall brand atmosphere.

\subsubsection{System Prompt}

With the brand inputs, we use an LLM to generate a shot plan. The system prompt instructs the LLM to adopt the persona of a creative director planning the shots in a short brand video. It has sections that define cinematic ads, shot types, camera movements, editing techniques, and pacing.

The system prompt first outlines the characteristics of cinematic ads: they evoke an emotional response using narration or music, use cinematic techniques, and have no traditional problem-solving narrative arc. The prompt explicitly states that cinematic ads are not problem-solving narratives, product demos, UGC, or slideshows.

The system prompt then lists common shot types (close-up, medium shot, and wide shot) and camera movements (e.g., no camera movement, pan, tilt, and push-in). We gathered the list from standard cinematography resources (e.g.,~\cite{studiobinder2020camera}). Although shot types and camera movements are production rather than editing choices, they shape shot selection during editing. We included shot types and camera movements in the system prompt so the LLM can use the vocabulary to describe a shot.

Next, the system prompt specifies a list of editing techniques with descriptions (e.g., montage, match cut, overlay-imagery alignment, cut on action, and cut to beat). We curated this list from Study 1b. We included an editing technique if it appears in any videos in our corpus.

The system prompt also encodes the pacing guidance: the average duration of a shot is 1.2 seconds and per-shot durations should be between 0.5 and 7.5 seconds. These are the temporal properties we observed in Study 1b.

\begin{figure*}
  \includegraphics[width=\textwidth]{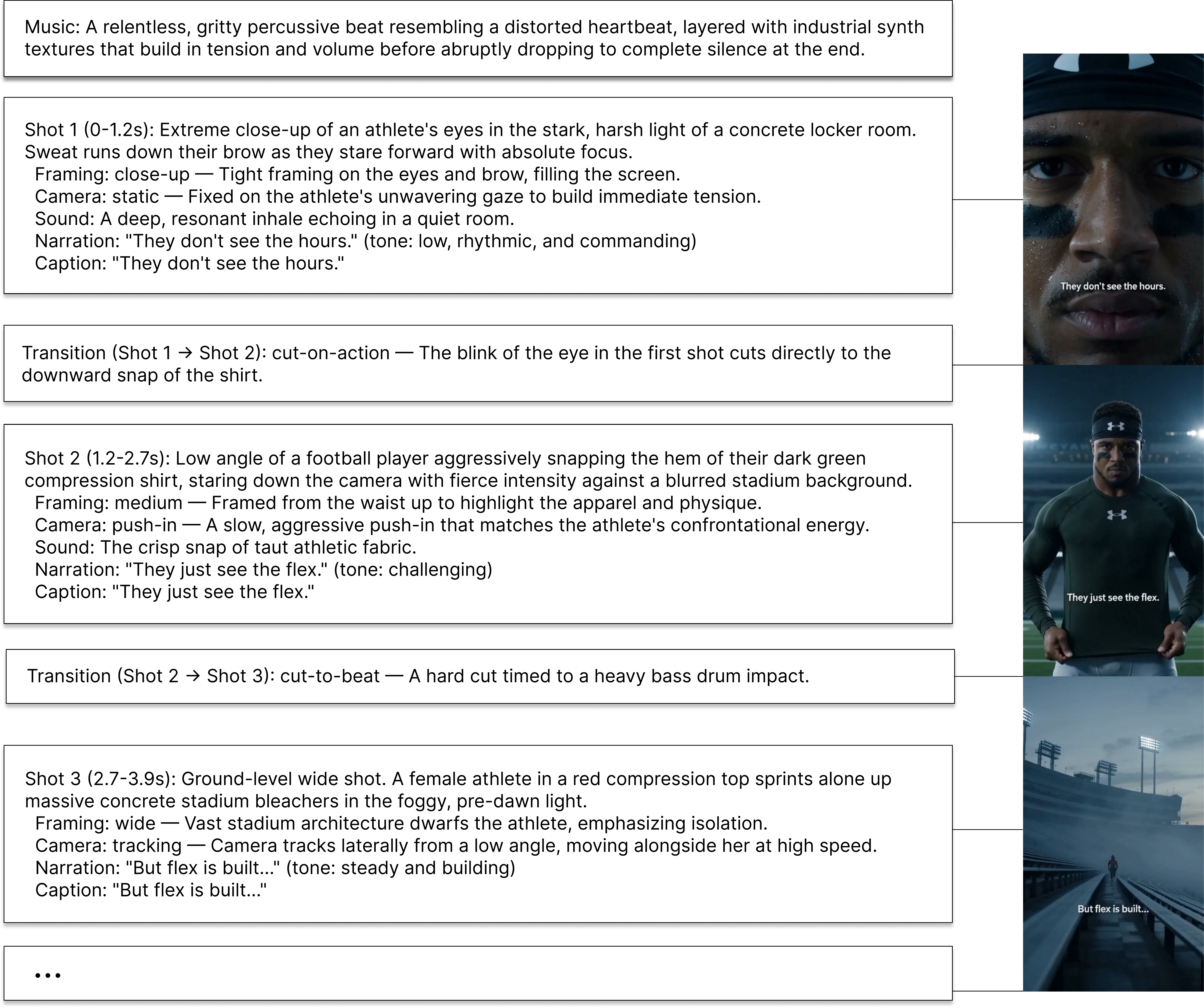}
  \caption{The first three shots in a shot plan for an Under Armour-inspired cinematic ad. The shot plan describes the music, the sequence of shots, and the transitions between shots. Frames from the first three shots of the resulting AI-generated video are shown on the right.}
  \Description{A shot-plan example for an Under Armour-inspired ad. The plan specifies percussive music and presents three numbered shots with their duration, framing, action, camera movement, narration, captions, and transitions. The shots depict an extreme close-up of an athlete's eyes, a football player pulling a compression shirt, and a woman sprinting in a stadium. Corresponding generated video frames appear beside the textual descriptions.}
  \label{fig:plan}
\end{figure*}

\subsubsection{Shot Plan}

The shot plan specifies the ad's music, the sequence of shots, and the transitions between shots. Figure~\ref{fig:plan} shows the first three shots in the shot plan for an Under Armour-inspired cinematic ad\footnote{Under Armour-inspired cinematic ad: \url{https://youtu.be/rmEU0QCFUh8}}.

The plan describes the music as \textit{``a percussive beat resembling a distorted heartbeat.''}

Shot 1 lasts 1.2s. It shows an extreme close-up of an athlete's eyes with no camera movement. Both narration and caption read, \textit{``They don’t see the hours.''} Shot 2 lasts 1.5s and shows \textit{``a football player snapping the hem of their dark green compression shirt.''} It is a medium shot where the camera slowly pushes in. The narration and caption both read, \textit{``They just see the flex.''} Shot 3 is a wide shot of a female athlete sprinting in a stadium.

The plan also specifies the transitions between shots. Shot 1 transitions to Shot 2 as the athlete in Shot 1 blinks his eyes. Shot 2 cuts to Shot 3 with the beat. The plan describes that we should hear \textit{``a heavy bass drum impact''} between the two shots.

\subsection{Generating Video}

With the shot plan, a video generation model (e.g., Seedance 2.0) renders the cinematic ad. We also pass the brand logo to the model so it can render the logo accurately. However, we do not pass the style reference images to the model. We found that passing these images caused the model to reproduce the reference frames too closely, yielding outputs that resembled existing advertisements rather than new cinematic ads with original editing.

As the Under Armour-inspired example illustrates, AI-generated cinematic ads can appear highly realistic, coherent, and cinematic. Yet, such high-level impressions do not explain which editing choices work well and which can be improved. To identify the dimensions to critique the editing quality of AI-generated cinematic ads, Study 2 asks professional video editors to evaluate these ads. 

\section{Study 2: Evaluating AI-Generated Cinematic Ads}

What dimensions do professional video editors use when evaluating the editing quality of AI-generated cinematic ads? In Study 2, we recruited professional editors to critique these ads.

\subsection{Methodology}

\subsubsection{Data Collection}

We generated cinematic ads using the web interface described in the previous section. Study 1a curated ads for 15 brands while Study 1b curated ads for 20 brands. For each of the 35 brands, we collected a textual description, the logo, a color palette, and style-reference images as inputs to the web interface. All 35 brands are real. We used real brands to make the evaluation tasks more realistic. For example, when editors recognize a brand, they can comment on whether an ad feels appropriate for the brand image.

An LLM generated the brand descriptions, which were manually checked by the research team. We collected the brand logos and color palettes from Brandfetch~\cite{brandfetch}, which is a data platform with publicly available brand materials. We collected three style-reference images for each brand by extracting frames from the ads collected in Studies 1a and 1b. These reference images are intended to provide the planner LLM with a brand’s visual qualities and inspirations when generating the shot plan.

With each brand's inputs, the web interface first generated a shot plan and then the video ad. For shot plan generation, we used two LLMs (Claude Opus 4.7 and Gemini 3 Pro) to introduce variations in the planning style. For video generation, we selected Seedance 2.0. We tested other video generation models, including Veo 3.1 and Kling 3.0. Other models, however, frequently dropped multiple shots from the plan, resulting in ads that felt incomplete. We removed these alternatives to prevent participants from disproportionately focusing on video incompleteness rather than the editing quality. Additionally, all generated videos were standardized to a 15-second duration, which Study 1b identified as the most frequent length in our cinematic ads corpus.

With 35 brands, 2 LLMs for shot plan generation, and 1 video generation model, we generated 35 x 2 x 1 = 70 video ads.

\subsubsection{Participants}

We recruited six professional video editors through Upwork. Participants were eligible if they had at least five years of experience editing professional brand, commercial, or social video content and were comfortable writing in English about editing craft using timestamps and editing terminology. Prior experience with AI-generated video was not required.

Participants reported between five and thirteen years of professional editing experience. Two participants evaluated seven ads each, and four participants evaluated fourteen ads each. We paid \$50 for evaluating every seven ads. After completing their assigned evaluations, we conducted a one-hour interview with each participant. We paid them an additional \$50 for the interview.

\subsubsection{Procedure}

Participants evaluated ads through a web interface. After providing informed consent, they viewed their assigned ads one at a time. For each ad, they answered four open-ended questions.

\begin{enumerate}
\item What editing choices work well?
\item What editing choices feel weak, ineffective, or missing?
\item How would you edit this differently?
\item Optional: Do you have any additional comments about this advertisement beyond its editing?
\end{enumerate}

We designed the first three questions to elicit professional judgment about effective editing choices, editing insufficiencies, and repair strategies. During a pilot study, we observed that participants sometimes mixed comments about AI-generation artifacts (e.g., awkward physics) into the first three questions. We therefore added the fourth question to give participants a separate place to discuss issues beyond editing. This nudged participants to focus Questions 1–3 on editing quality while allowing them to document generation defects when those defects affected their experience of the video.

After participants completed their assigned critiques, we conducted a one-hour interview with each participant. The written responses may refer to an audio effect or visual moment that could be difficult to interpret from the text alone. During an interview, the researcher went over the ads one by one. The participant elaborated and disambiguated their written responses.

\subsubsection{Analysis}

We segmented the written responses into 870 sentences. An author open-coded the sentences to develop an initial codebook. We provide the codebook in the supplementary materials. Two coders then independently coded a randomly selected 25\% of the data using the codebook. They discussed disagreements and refined the codebook. We repeated this process with additional randomly selected 25\% subsets until Cohen's Kappa was above 0.7. After the codebook was finalized, one coder coded the remaining sentences.

\subsection{Dimensions of Editing Quality in Cinematic Ads}

Our analysis identified six dimensions that professional editors used to evaluate the editing quality of AI-generated cinematic ads. For each dimension, we provide the definition in a question form and the frequency. We further provide examples from participants' critiques and group these examples into categories.

\subsubsection{Narrative Progression}

Does the narrative develop coherently from beginning to end? Five participants discussed this dimension in 139 sentences across 46 videos. Participants commented on narrative coherence, narrative ordering, opening, and closing.

\paragraph{Narrative Coherence}

Do the shots feel like they belong to the same story?

For instance, in a Sephora-inspired ad\footnote{Sephora-inspired cinematic ad: \url{https://youtu.be/6_MdWlTd950}}, a shot shows a rapid macro of a lipstick bullet being twisted up, followed by a close-up of glossed lips. This sequence effectively implied using the product on the lips. P4 considered the association to be effective even though the model failed to match the exact shades.

\begin{figure}
  \includegraphics[width=\columnwidth]{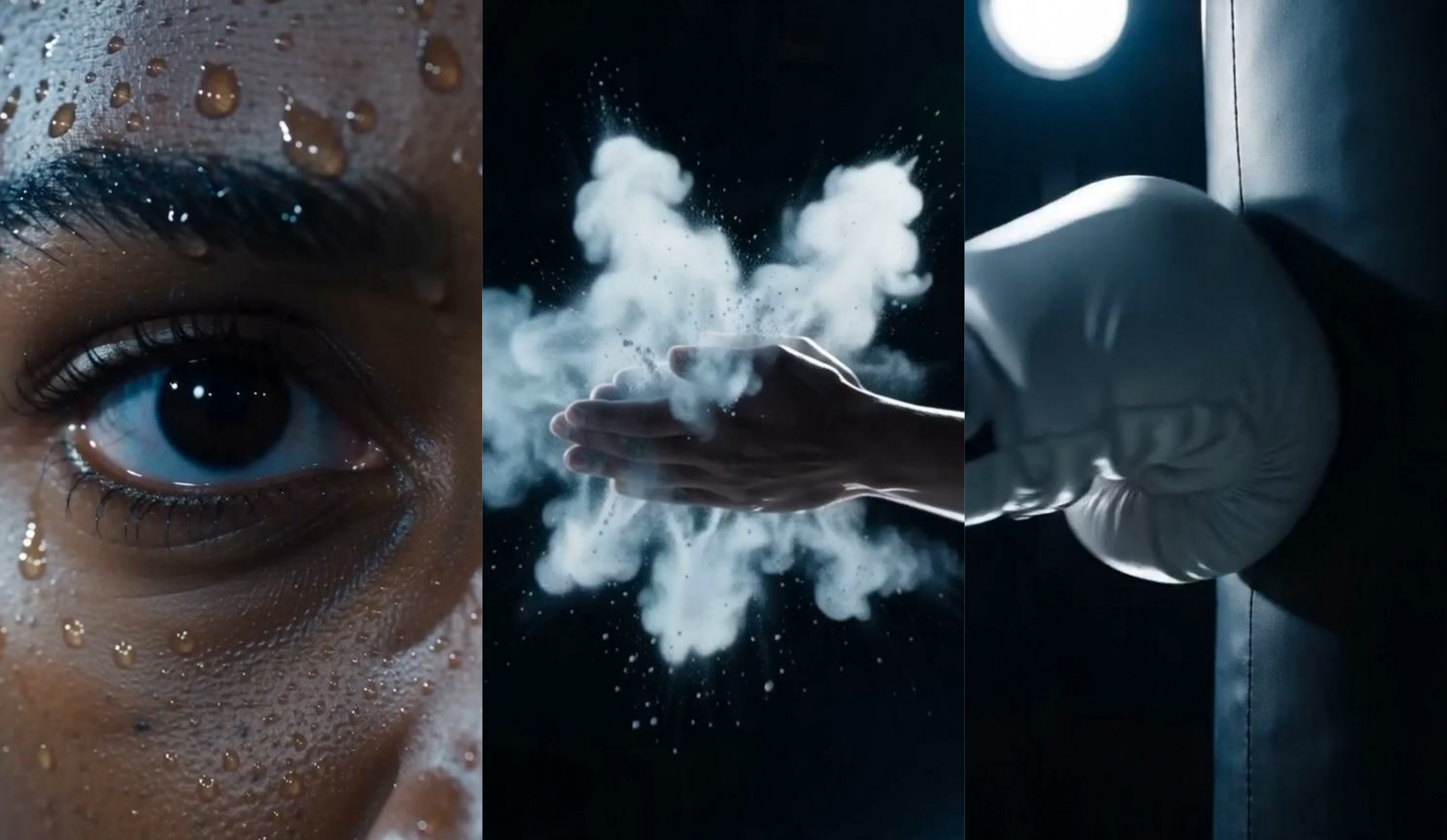}
  \caption{A Gatorade-inspired ad presenting a sprinter's eyes, chalk being slapped, and a punching boxer. P5 argued that the chalk shot seemed unrelated to sprinting and boxing.}
  \Description{Three frames from a Gatorade-inspired ad show a close-up of a sprinter's eyes, a cloud of chalk produced by hands striking together, and a boxer throwing a punch. The middle image has no clear visual connection to the athletes in the surrounding images.}
  \label{fig:coherence-gatorade}
\end{figure}

Occasionally, the videos included filler shots that did not fit the story. A Gatorade-inspired ad\footnote{Gatorade-inspired cinematic ad: \url{https://youtu.be/a2Bg5aM53Nw}} presented a sequence of a sprinter's eyes, chalk being slapped, a sprinting soccer player, and a punching boxer (Figure~\ref{fig:coherence-gatorade}). P5 argued that the chalk shot felt out of place: \textit{``the chalk seems to be unrelated to the running [...] and also unrelated to the boxing shot [...] I'm left thinking, why is someone randomly clapping chalk into the air?''} They suggested starting \textit{``the scene with a push in of a woman's eye. That woman should be a gymnast, and then the chalk scene, then a shot of a gymnast on bars.''} They believed \textit{``that would tell a more cohesive story.''}

\paragraph{Narrative Ordering}

Are the shots arranged in an order that makes the narrative understandable and reasonable?

\begin{figure}
  \includegraphics[width=\columnwidth]{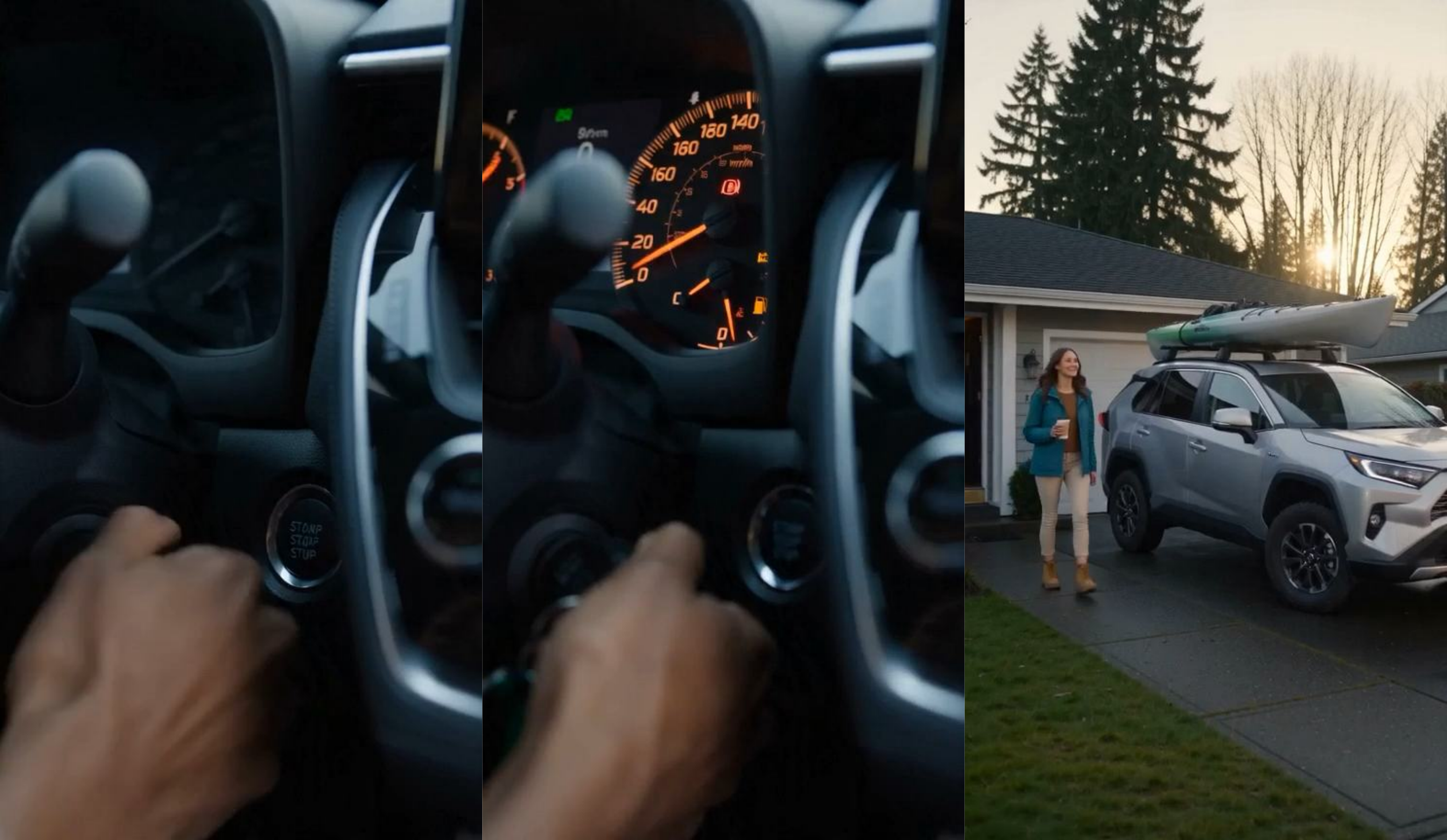}
  \caption{A Toyota-inspired ad displaying a key turning and then a woman approaching the car. The sequence could be reversed.}
  \Description{Frames from a Toyota-inspired ad show a car key being turned before a woman is shown approaching the car, creating a sequence in which the car appears to be started before the driver reaches it.}
  \label{fig:order-toyota}
\end{figure}

A Toyota-inspired ad\footnote{Toyota-inspired cinematic ad: \url{https://youtu.be/AL7TUpfHZPY}} first displayed a close-up of a key turning before cutting to a woman approaching the car (Figure~\ref{fig:order-toyota}). Identifying this sequencing error, P2 said that they would swap the shots: \textit{``I would take the shot of the woman walking to the car [...] and move it before the closeup of the key turning in the engine.''}

\begin{figure}
  \includegraphics[width=\columnwidth]{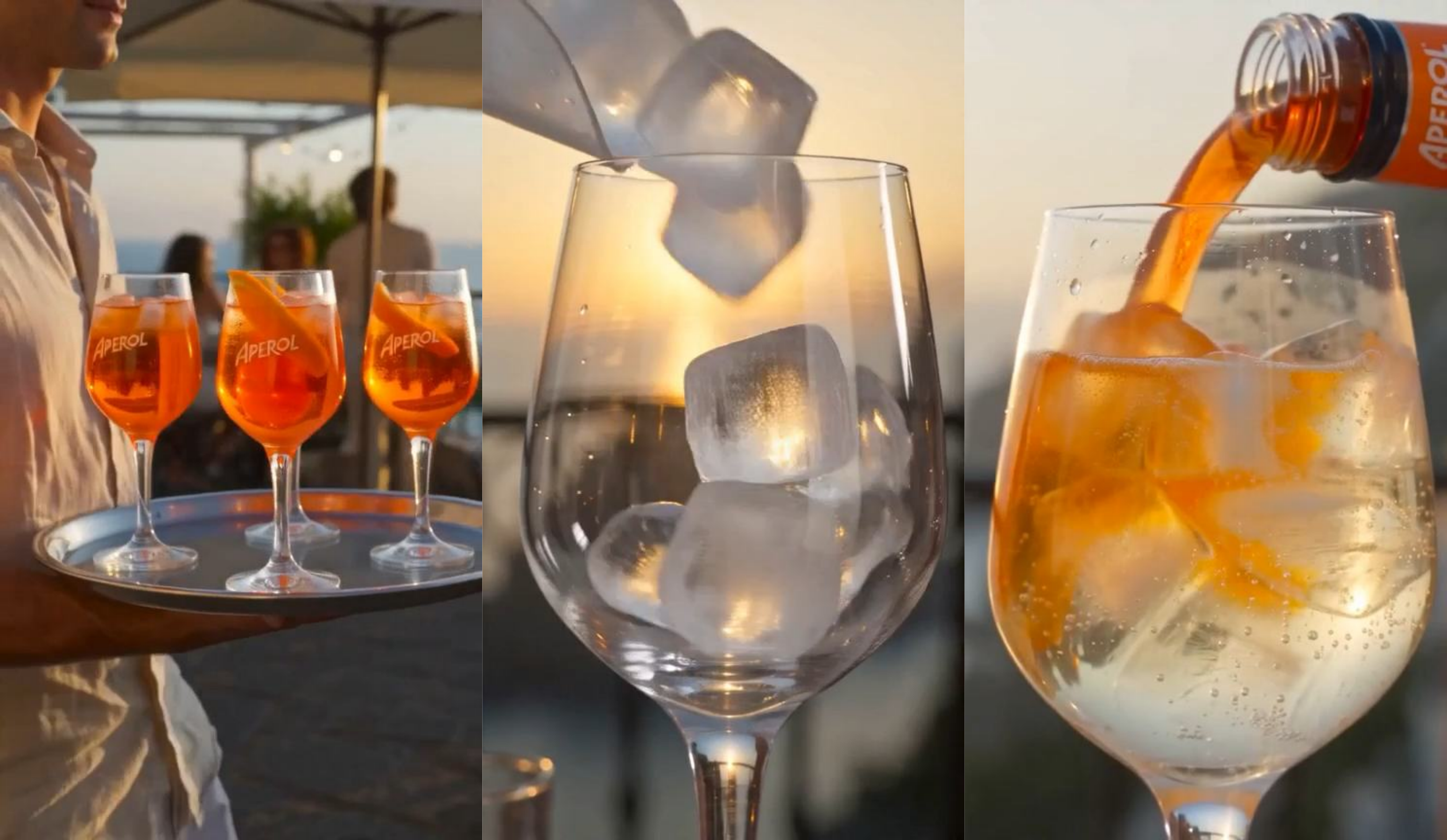}
  \caption{An Aperol-inspired ad presenting a waiter delivering drinks before the drinks being prepared. The sequence could be reversed.}
  \Description{Frames from an Aperol-inspired ad first show a waiter delivering finished drinks and then show the drinks being prepared, reversing the expected chronological order.}
  \label{fig:order-aperol}
\end{figure}

Similarly, an Aperol-inspired ad\footnote{Aperol-inspired cinematic ad: \url{https://youtu.be/eKJqJTd4myY}} presented a waiter delivering drinks prior to a shot of the drinks being prepared (Figure~\ref{fig:order-aperol}). P4 wanted to reverse the order, explaining that \textit{``showing the server walking with the Aperol drinks, then cutting back to showing them being made is a weird timeline. I would start with [...]  the scenery, making the drink, the waiter walking, handing the drinks, people enjoying the drink in the scenery to tell a cohesive story.''}

\paragraph{Opening}

Is the opening effective (e.g., using an establishing shot to provide context or using a hook to draw attention)?

\begin{figure}
  \includegraphics[width=\columnwidth]{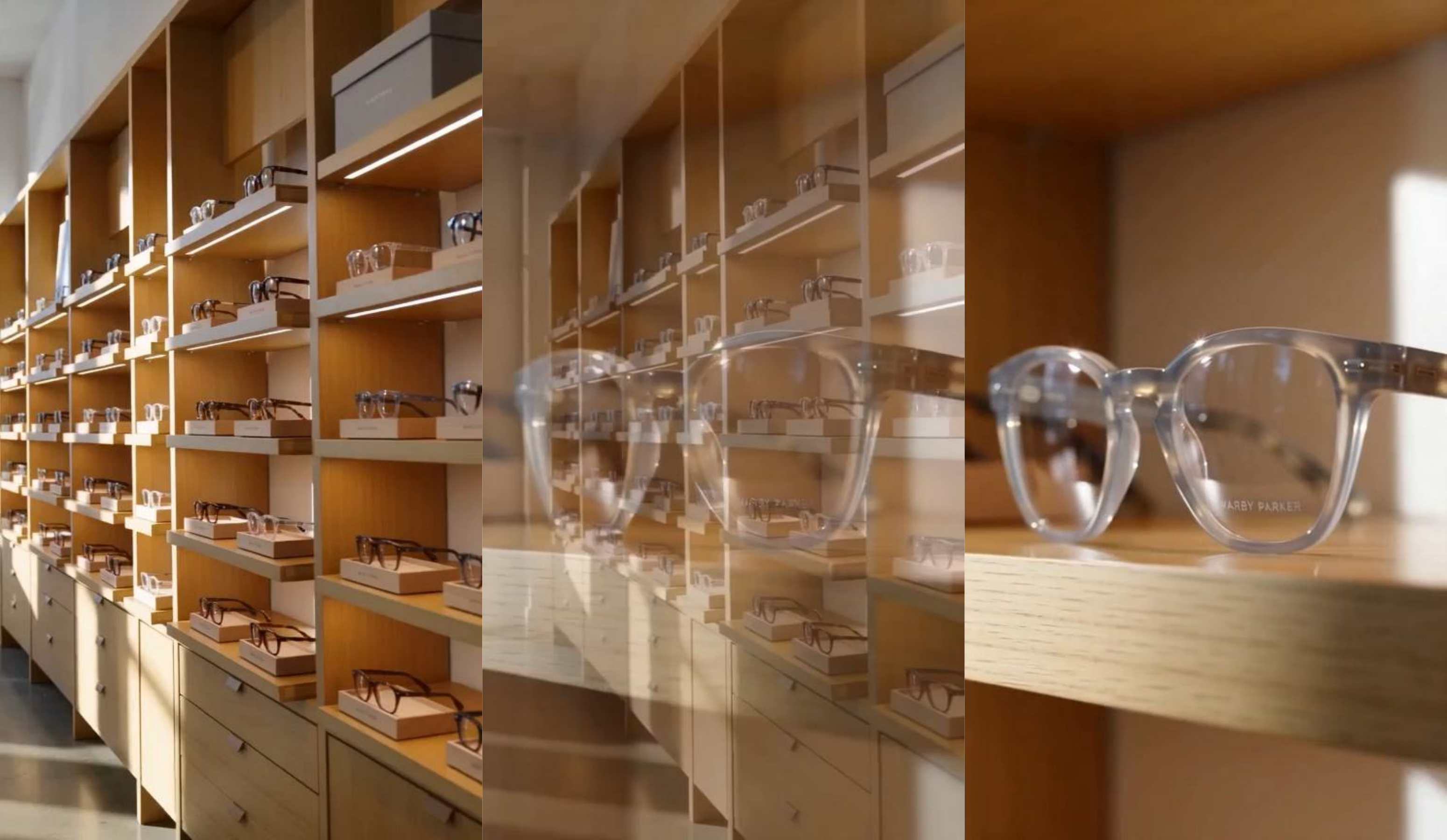}
  \caption{A Warby Parker-inspired ad shows a showroom with rows of eyeglasses and then a pair of glasses. The showroom opening shot effectively establishes context for the ad.}
  \Description{The opening frames of a Warby Parker-inspired ad move from a wide view of a showroom containing rows of eyeglasses to a closer view of an individual pair of glasses.}
  \label{fig:open-warby}
\end{figure}

For example, a Warby Parker-inspired ad\footnote{Warby Parker-inspired cinematic ad: \url{https://youtu.be/E4uHMDiMD8I}} uses a wide shot of a showroom interior with rows of eyeglasses as an establishing shot and then shows a close-up of a single pair of glasses (Figure~\ref{fig:open-warby}). P5 noted, \textit{``I like the establishment shots of all the glasses in the store, and the close up shot of the glasses.''}

\begin{figure}
  \includegraphics[width=\columnwidth]{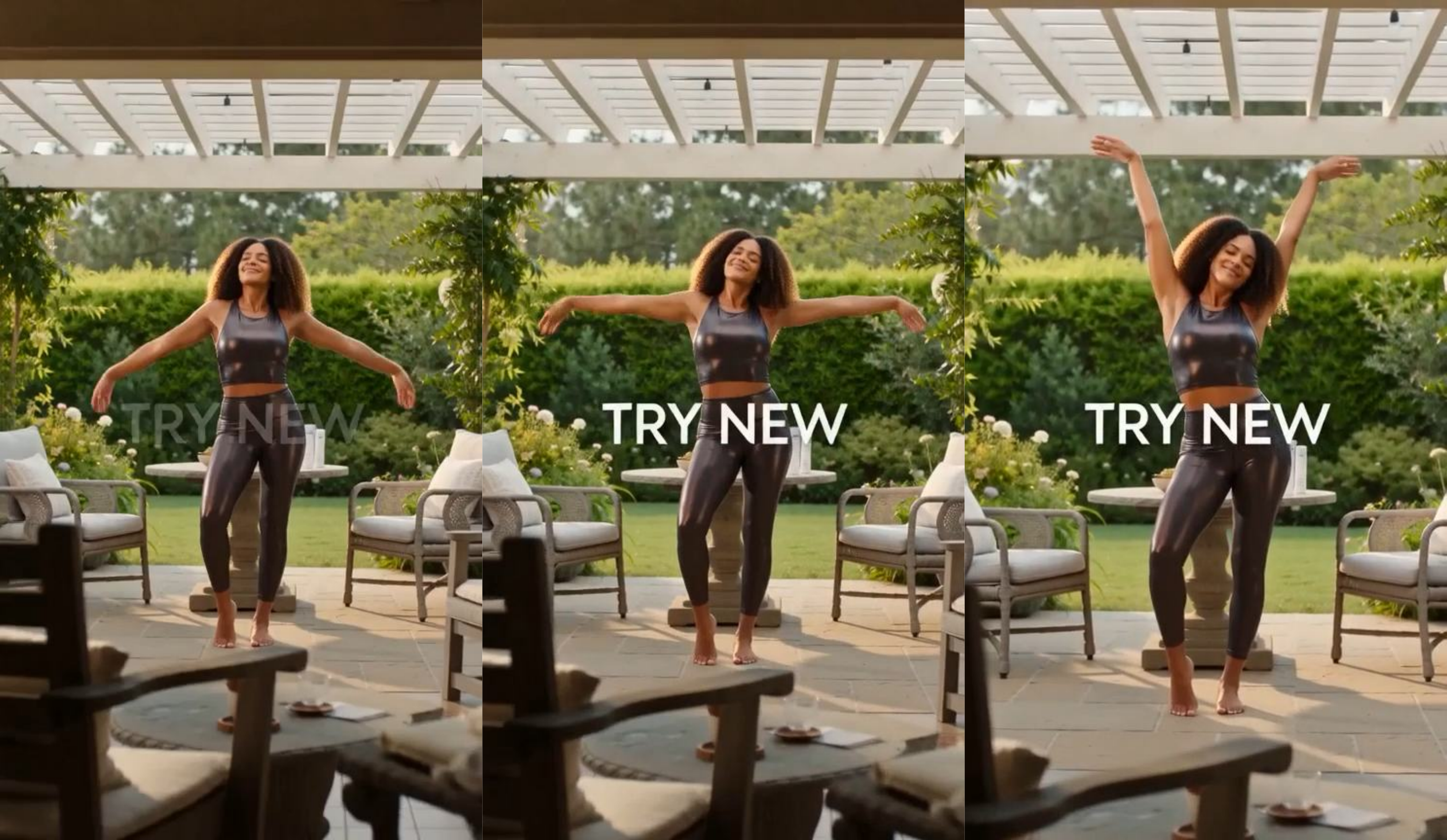}
  \caption{A Dove-inspired ad showing a woman stretching her arms joyfully with the voiceover saying \textit{``this summer''}. P3 praised how the woman's joy captures viewers' curiosity.}
  \Description{Opening frames from a Dove-inspired ad show a smiling woman outdoors stretching both arms upward. Her joyful movement is presented against bright summer scenery.}
  \label{fig:open-dove}
\end{figure}

Other participants commented on how an ad opened with an effective hook. In a Dove-inspired ad\footnote{Dove-inspired cinematic ad: \url{https://youtu.be/hDS_fhh9NQU}}, the first shot shows a woman stretching her arms joyfully. The camera pushes in and the voiceover says, \textit{``this summer''} (Figure~\ref{fig:open-dove}). P3 praised the opening: \textit{``the opening push-in paired with the on-screen text builds curiosity and pulls attention right away.''}

\paragraph{Closing}

Does the closing feel abrupt or smooth?

In a Chipotle-inspired ad\footnote{Chipotle-inspired cinematic ad: \url{https://youtu.be/WcdElS9w1hs}}, the music suddenly cut off toward the end. P6 argued that \textit{``the weird music cut off at second 12 does not fit''} and that \textit{``it felt like it was a mistake more so than an artistic choice.''}

\begin{figure}
  \includegraphics[width=\columnwidth]{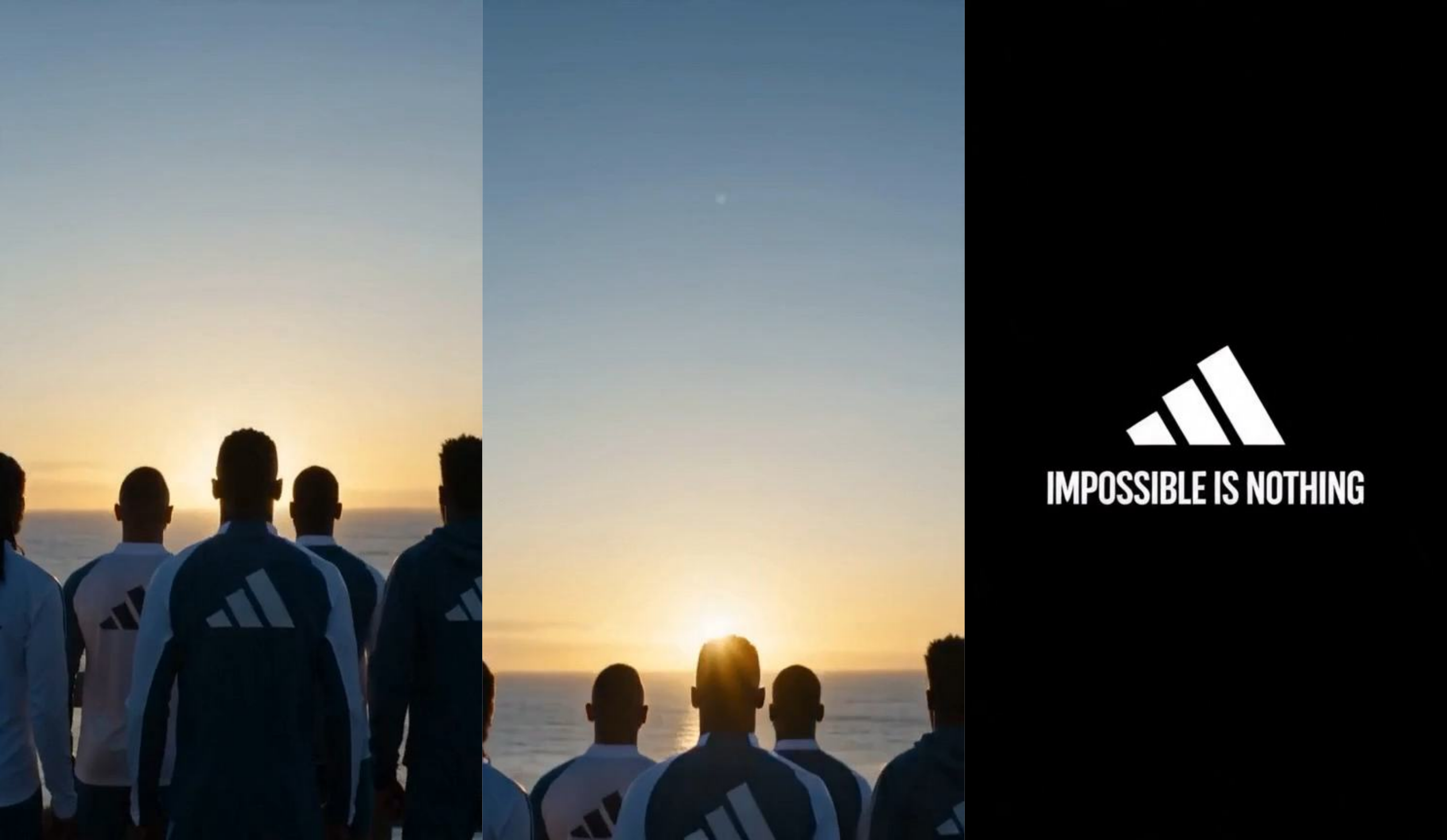}
  \caption{An Adidas-inspired ad ending with a group of runners looking at the sky as the camera pans up. P1 commented that the logo could be directly projected onto the sky and not shown on a black screen (frame 3).}
  \Description{Closing frames from an Adidas-inspired ad show a group of runners looking upward as the camera moves toward the sky, followed by an Adidas logo displayed separately on a black background.}
  \label{fig:close-adidas}
\end{figure}

An Adidas-inspired ad\footnote{Adidas-inspired cinematic ad: \url{https://youtu.be/Yl7s5DH5AHI}} ends with a group of runners looking out over the sunrise horizon as the camera pans up to the sky. It cuts to a static black screen with the logo (Figure~\ref{fig:close-adidas}). P1 commented that the transition felt \textit{``abrupt.''} They suggested a more integrated finish where the Adidas logo would \textit{`` fade directly into the sky plate itself.''}

\subsubsection{Audiovisual Coordination and Sound Design}

Do sound, music, silence, and visual editing work together to support the rhythm, meaning, and emotional effect of the ad? All six participants discussed this dimension in 135 sentences across 51 videos. Participants commented on beat synchronization, sound-action matching, sound bridges, musical progression, sound variation, and silence.

Since it is hard to depict audio quality in text, to facilitate understanding, we provide timestamps when discussing example critiques from participants.

\paragraph{Beat Synchronization}

Is cut to beat used effectively?

\begin{figure}
  \includegraphics[width=\columnwidth]{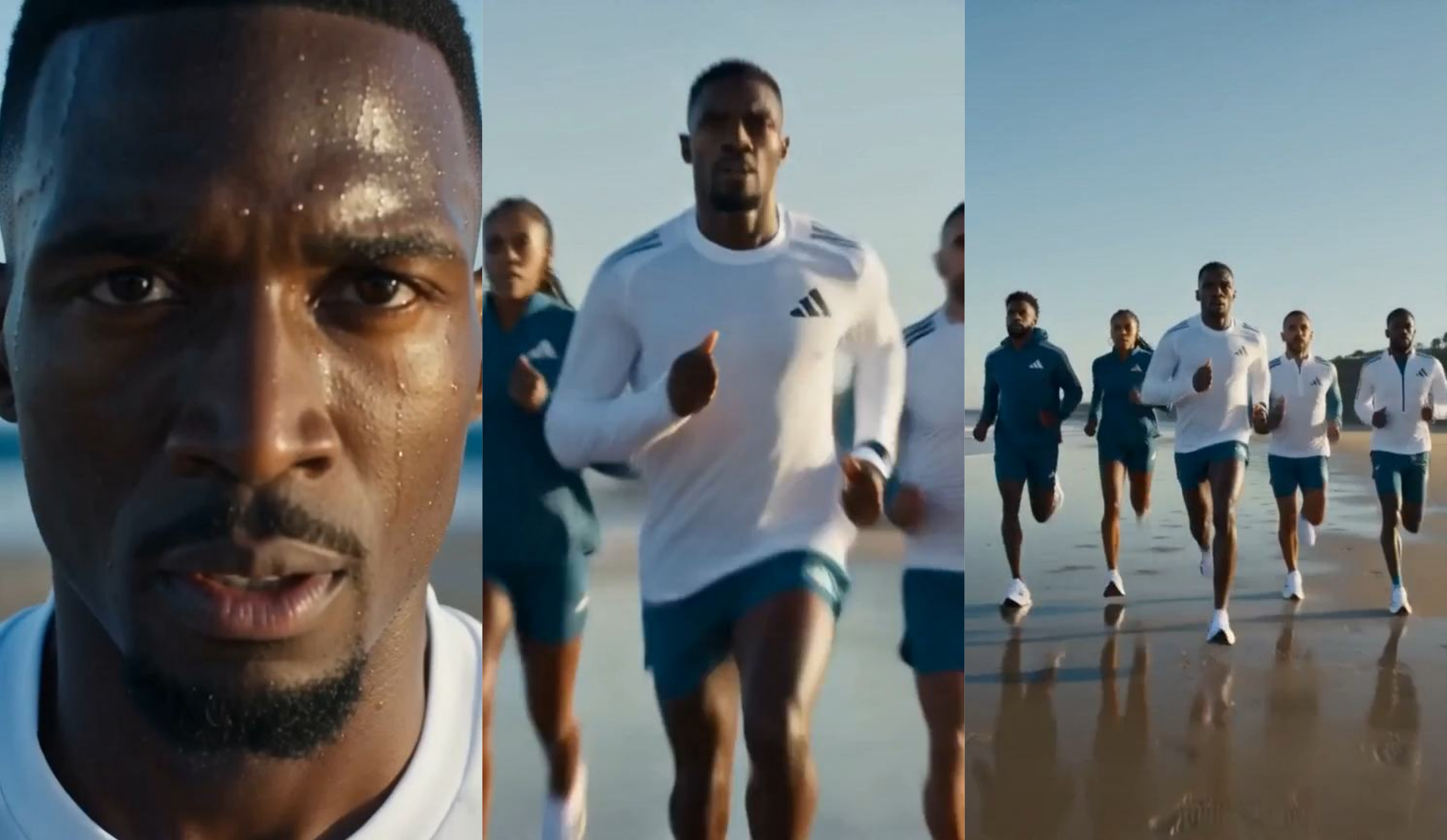}
  \caption{An Adidas-inspired ad showing a close-up of a runner's face synchronized with a beat drop (frame 1) and then a group of runners (frames 2 - 3). P1 commented on the effectiveness of the beat drop cut.}
  \Description{Three frames from an Adidas-inspired ad. The first is a tight close-up of a runner's face at the moment of a beat drop. The following frames widen to show a group of runners moving together.}
  \label{fig:beat-adidas}
\end{figure}

An Adidas-inspired ad\footnote{Adidas-inspired cinematic ad: \url{https://youtu.be/Yl7s5DH5AHI}} has a close-up of a runner's face synchronized with a beat drop (0:05). It then cuts to a wide shot of a group of runners (Figure~\ref{fig:beat-adidas}). P1 felt that the beat drop cut was impactful: \textit{``The cut from the extreme close-up of the runner's face to a wide shot of a full crew running toward the camera is timed perfectly with a heavy, satisfying bass drop. It shifts the narrative from solitary struggle to collective power.''}

\paragraph{Sound-Action Matching}

Do the sound effects and Foley correspond convincingly to the action shown on screen?

P4 complimented the bottle-opening sound effects in a Heineken-inspired ad\footnote{Heineken-inspired cinematic ad: \url{https://youtu.be/3LhzH27kzro}} (0:01) while P1 liked the \textit{``crunch of a sneaker hitting the pavement''} (0:02) and \textit{``the natural hum of the wind blowing past''} (0:08) in a Lululemon-inspired ad\footnote{Lululemon-inspired cinematic ad: \url{https://youtu.be/E0B7Kp-zt74}}.

\paragraph{Sound Bridges}

Do J-cuts, L-cuts, and continuing sound effects help connect shots effectively?

An Under Armour-inspired ad\footnote{Under Armour-inspired cinematic ad: \url{https://youtu.be/OQk2AbQ9SDg}} has a wide shot of a basketball player shooting (0:08), followed by a close-up of a girl wrapping athletic tape around her wrist. P2 expected \textit{``a swoosh sound for the basketball shot at 00:08 with an L cut going into the next clip.''} Without this sound, P2 said the shot of a basketball player \textit{``feels incomplete.''}

An Alo Yoga-inspired ad\footnote{Alo Yoga-inspired cinematic ad: \url{https://youtu.be/WImCoQ9OH_s}} has constant breath sounds (e.g., 0:05) and dissolve transitions (e.g., 0:01). P2 suggested that a breath sound should overlap with a dissolve transition and that this \textit{``would make the piece feel a bit more serene and dreamy.''}

\paragraph{Musical Progression}

Does the music effectively develop the energy to support the narrative progression of the ad?

For example, when commenting on a Sephora-inspired ad\footnote{Sephora-inspired cinematic ad: \url{https://youtu.be/6_MdWlTd950}}, P3 said that the music between 0:06 and 0:08 swells with the shots and movement to build a climax, \textit{``giving the feeling of exhilaration and excitement.''}

In contrast, P1 noted that the music in a New Balance-inspired ad\footnote{New Balance-inspired cinematic ad: \url{https://youtu.be/0JEvOoI70xU}} seems flat: \textit{``The background beat is catchy and rhythmic, but it stays at roughly the same energy level throughout the entire 14 seconds. The edit misses a musical crescendo or an audio 'drop' to signify a climax before hitting the final logo screen.''}

\paragraph{Silence}

Is silence used intentionally and effectively to establish a hook, heighten focus, and signify a resolution?

\begin{figure}
  \includegraphics[width=\columnwidth]{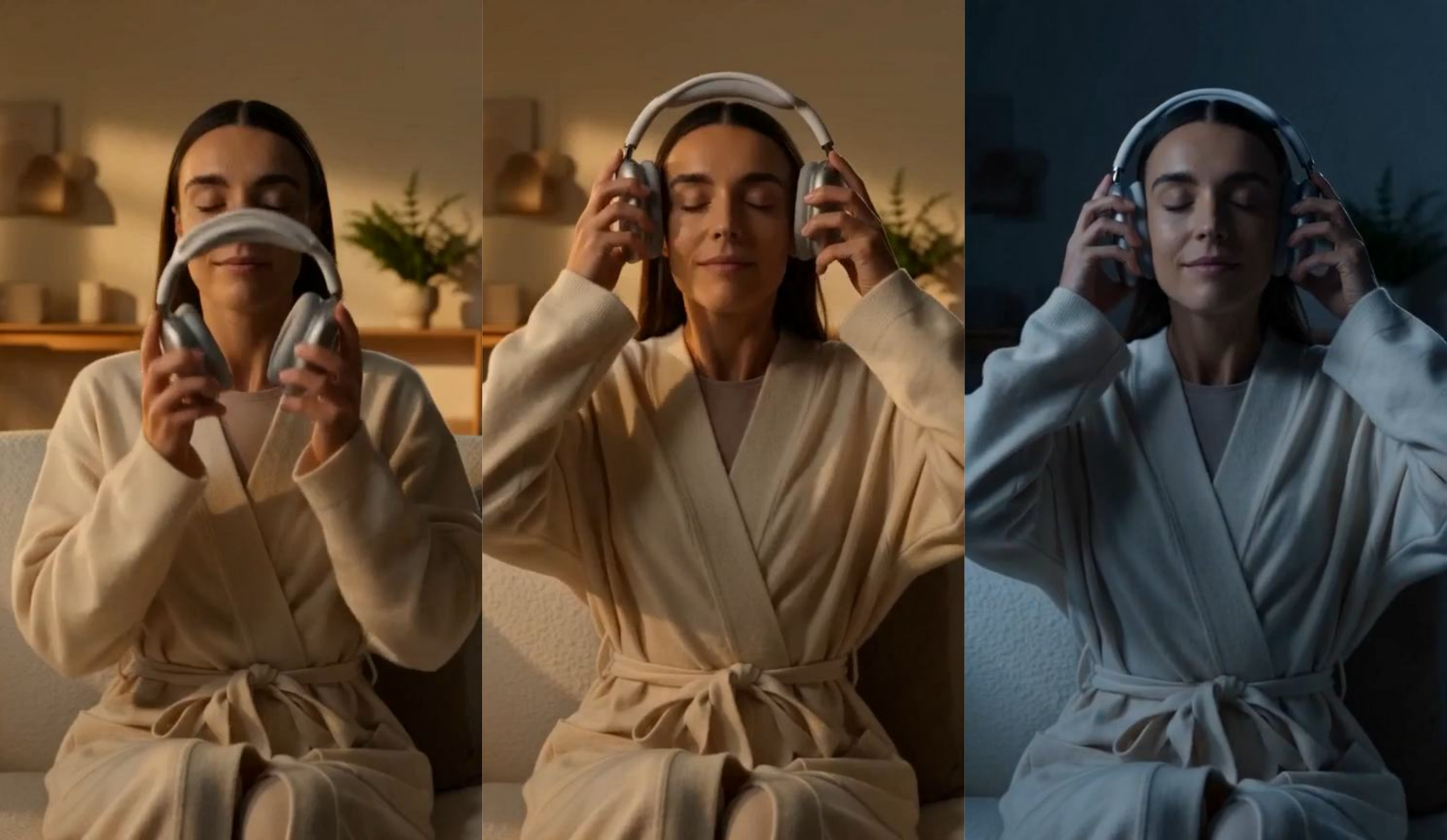}
  \caption{A Bose-inspired ad showing a woman putting on noise-cancelling headphones. Viewers hear silence as the warm background lighting shifts to a cool blue tone. The silence effectively captures viewers' focus.}
  \Description{Frames from a Bose-inspired ad show a woman placing noise-cancelling headphones over her ears. As she puts them on, the background lighting changes from a warm orange tone to a cool blue tone.}
  \label{fig:silence-bose}
\end{figure}

For example, a Bose-inspired ad\footnote{Bose-inspired cinematic ad: \url{https://youtu.be/i8VS8vZbJPs}} has a shot where a woman places a pair of noise-cancelling headphones over her ears, and the warm background lighting instantly shifts to a cool blue tone (Figure~\ref{fig:silence-bose}). As the woman puts on the headphones, we hear silence (0:11). P3 praised the silence, commenting that it \textit{``makes viewers focus more.''}

Silence can sometimes be confusing. A Dior-inspired ad\footnote{Dior-inspired cinematic ad: \url{https://youtu.be/uFGQLCMDO50}} shows a close-up of a woman's lips in the first shot. This shot features a soft inhale. P4 commented that \textit{``the lack of SFX on the opening breath shot, is confusing. I had to check if I had my sound on.''}

\subsubsection{Visual Composition and Graphics}

Do the visual arrangement, framing, captions, color, and scene variation guide attention and make the ad feel polished? All six participants discussed this dimension in 114 sentences across 41 videos. Participants commented on framing, focal attention, caption design, color consistency, and scene variation.

\paragraph{Framing}

Are the objects framed appropriately for the intended visual emphasis and for mobile devices?

\begin{figure}
  \includegraphics[width=\columnwidth]{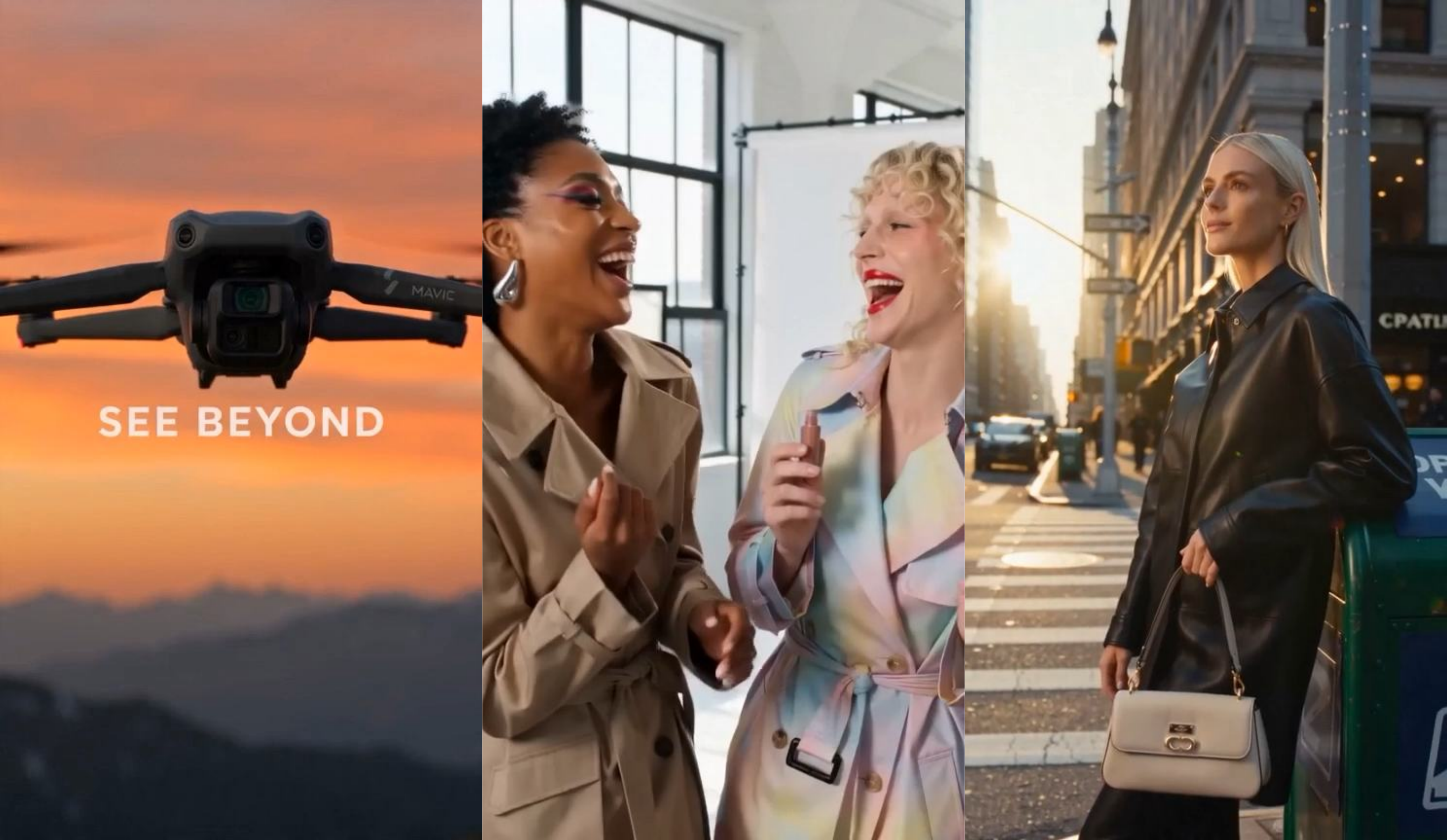}
  \caption{A DJI drone where the wings are cut off (frame 1), two women seated too close to the edges in a Sephora-inspired ad (frame 2), and a Coach bag at the bottom of the frame when viewers often look at the center of the frame (frame 3).}
  \Description{Three examples of framing problems. The first frame crops the outer portions of a DJI drone's wings. The second places two seated women close to the left and right edges of the frame. The third places a Coach bag near the bottom of the image, away from the visual center.}
  \label{fig:frame-dji-sephora-coach}
\end{figure}

A DJI-inspired ad\footnote{DJI-inspired cinematic ad: \url{https://youtu.be/XxU_dL_yxJw}} shows a drone with the wings cut off (Figure~\ref{fig:frame-dji-sephora-coach} frame 1). P2 said, \textit{``I would have used a slightly wider shot to show the entire drone in the final shot.''}

During the interview, P3 highlighted that vertical ads on Instagram and TikTok are often cropped off on new iPhones. They emphasized that the main subjects should be within some safe margins and not too close to the edges. For instance, in a Sephora-inspired ad\footnote{Sephora-inspired cinematic ad: \url{https://youtu.be/FY9I_vcHibQ}}, two women are seated close to the edges of the frame and might be cropped on some devices (Figure~\ref{fig:frame-dji-sephora-coach} frame 2): \textit{``I'd keep all key elements within the safe margins''} (P3).

\paragraph{Focal Attention}

Does the composition guide the viewer’s eye toward the intended object?

In cinematography, eye trace is a technique to anticipate where viewers look and position the element to meet that gaze~\cite{smith2012attentional}. For instance, a Coach-inspired ad\footnote{Coach-inspired cinematic ad: \url{https://youtu.be/tuHGfm9N1zU}} shows a Coach bag at the bottom of the screen (Figure~\ref{fig:frame-dji-sephora-coach} frame 3). Anticipating that viewers often look at the center, P5 noted, \textit{``I think the Coach bag could be more central in the final shot.''}

\begin{figure}
  \includegraphics[width=\columnwidth]{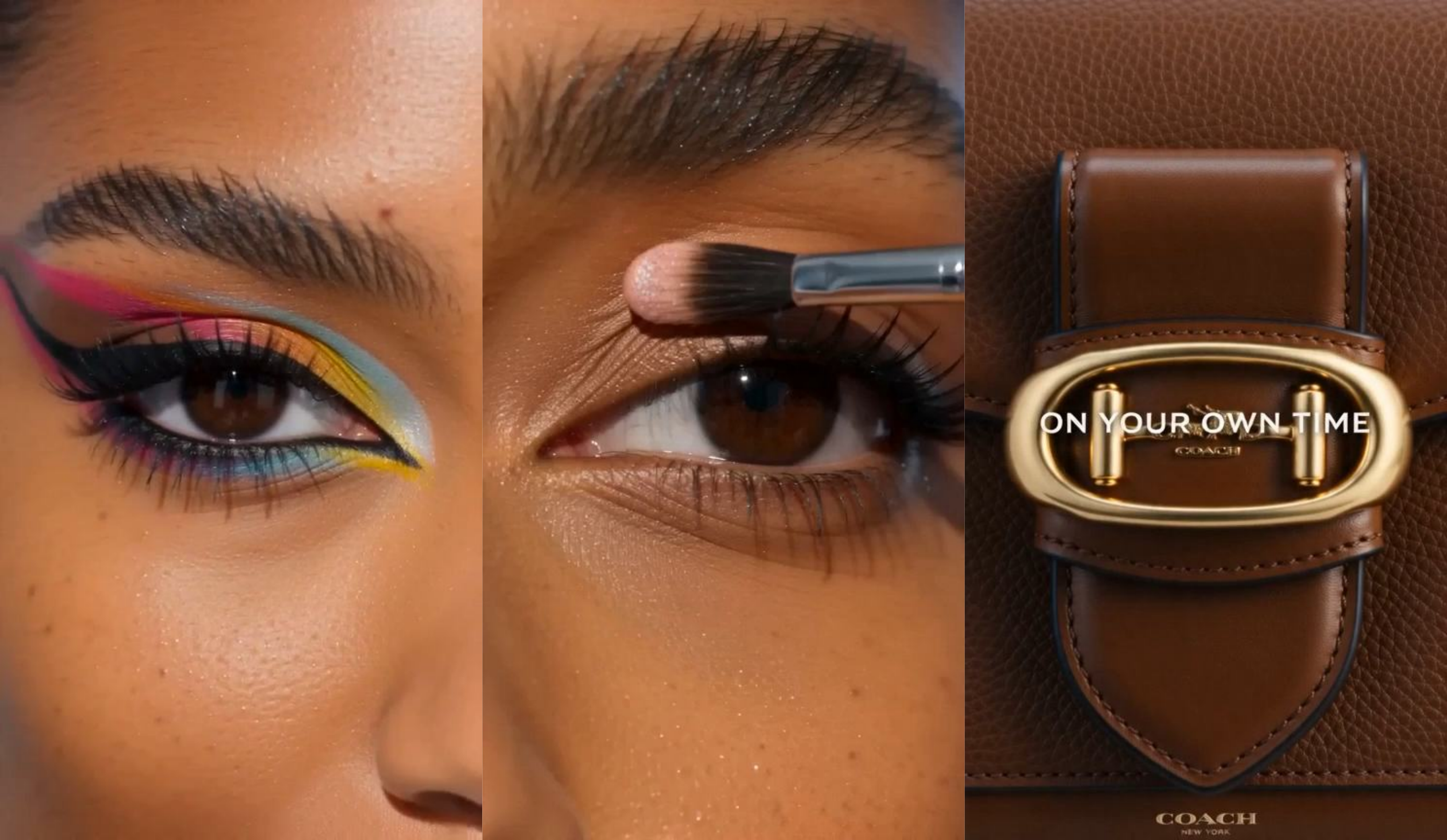}
  \caption{The pupil position shifts between consecutive shots in a Sephora-inspired ad (frames 1 - 2). The caption \textit{``ON YOUR OWN TIME''} is barely legible in a Coach-inspired ad (frame 3).}
  \Description{Three frames illustrate focal-point and text-legibility problems. In the first two frames, the position of a woman's pupil changes noticeably between adjacent close-ups. In the third, the small caption ``ON YOUR OWN TIME'' has little contrast with the background and is difficult to read.}
  \label{fig:focal-sephora-coach}
\end{figure}

Furthermore, maintaining eye trace is critical across two adjacent shots to prevent the viewer's gaze from jumping abruptly between cuts~\cite{murch2001blink}. A Sephora-inspired ad\footnote{Sephora-inspired cinematic ad: \url{https://youtu.be/FY9I_vcHibQ}} has two consecutive close-ups of an eye (Figure~\ref{fig:focal-sephora-coach} frames 1 - 2). Noticing that the pupil position shifted between shots, P3 said, \textit{``the position should match that of the other eye from the previous shot.''}

\paragraph{Caption Design}

Are the captions well placed, appropriately styled, and well integrated with the image?

In a Glossier-inspired ad\footnote{Glossier-inspired cinematic ad: \url{https://youtu.be/cHLZik16Gcc}}, the caption covers a model’s face (Figure~\ref{fig:motion-glossier} frame 3). P3 said, \textit{``the title text sits across the model's face — dropping it lower, below her features, would keep her face clear and read better.''}

Sometimes, the caption is not clearly legible since it does not contrast well with the background image. Figure~\ref{fig:focal-sephora-coach} frame 3 is an example from a Coach-inspired ad\footnote{Coach-inspired cinematic ad: \url{https://youtu.be/nzb9-BZh7fk}}. The caption \textit{``ON YOUR OWN TIME''} is difficult to read because the white font is overlaid on a golden brass clasp, and P6 noted, \textit{``you can't quite see the text.''}

\paragraph{Color Consistency} Does the color grading feel consistent and is it aligned with the brand’s image?

For example, a Mercedes-Benz-inspired ad\footnote{Mercedes-Benz-inspired cinematic ad: \url{https://youtu.be/Mc5XHxzh0rQ}} employs a dark blue color scheme and cyan ambient lighting. P3 said, \textit{``The video's colors are premium, which fits Mercedes.''}

\begin{figure}
  \includegraphics[width=\columnwidth]{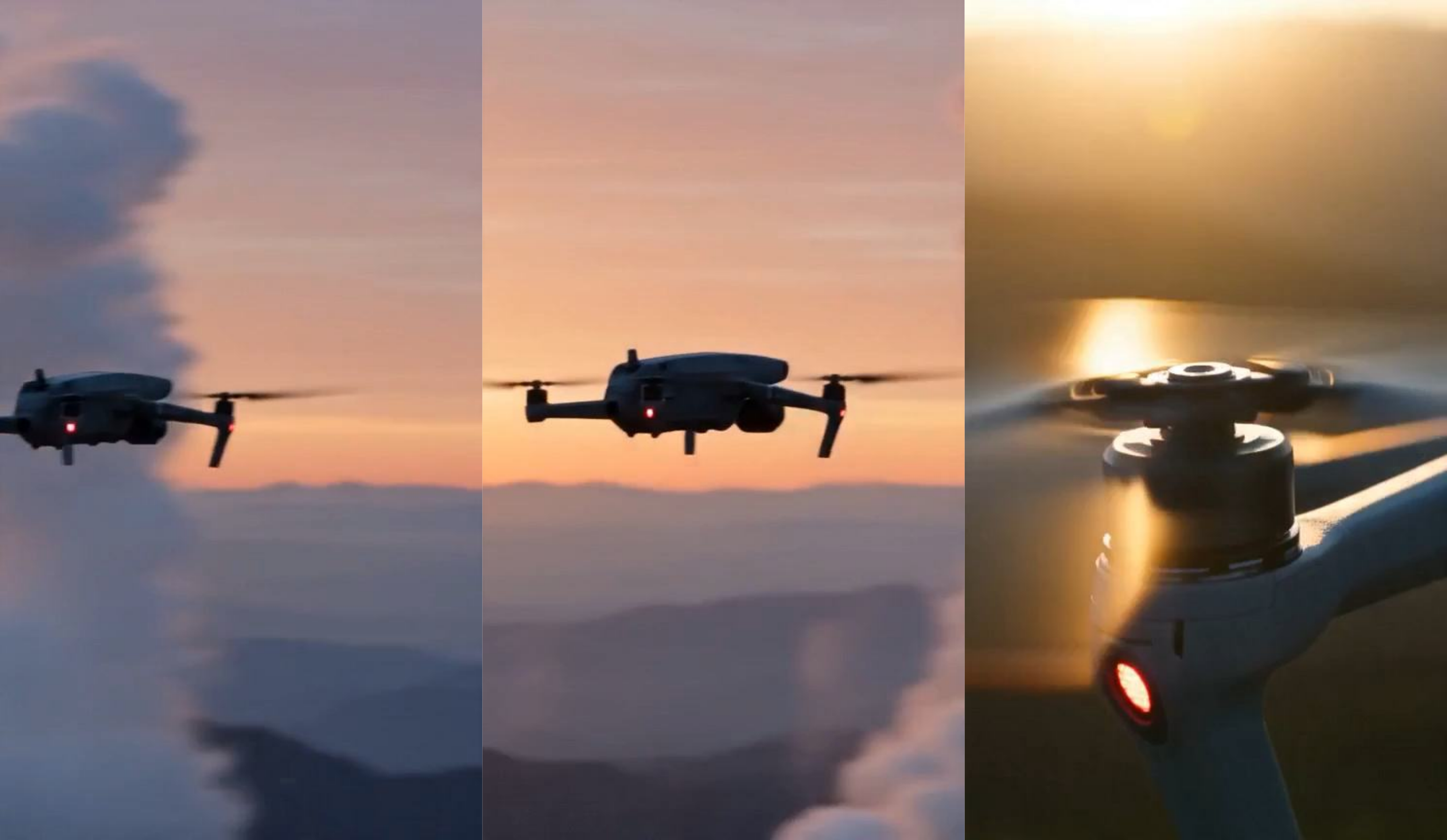}
  \caption{In a DJI-inspired ad, the first shot (frames 1 - 2) has a warmer golden hue while the second (frame 3) has a brighter golden color. Both shots capture the same sunset but the colors feel inconsistent.}
  \Description{Three frames from a DJI-inspired sunset sequence. The first two frames have a muted, warm golden hue, while the third frame depicts the same sunset using a noticeably brighter and more saturated golden color.}
  \label{fig:color-dji}
\end{figure}

During the interview, P3 observed that \textit{``the color grading isn't consistent across''} a DJI-inspired ad\footnote{DJI-inspired cinematic ad: \url{https://youtu.be/jj2x94Uu8h8}}. For instance, Figure~\ref{fig:color-dji} shows two consecutive shots depicting the same sunset scene. The first shot (Figure~\ref{fig:color-dji} frames 1 - 2) has a warmer golden hue while the second has a brighter golden color (Figure~\ref{fig:color-dji} frame 3). P3 pointed out that the shift in color temperature made the two shots feel like they were filmed using different cameras and that they should be color-corrected.

\paragraph{Scene Variation}

Does the ad have enough scene variation and are there sequences of scenes that feel repetitive?

\begin{figure}
  \includegraphics[width=\columnwidth]{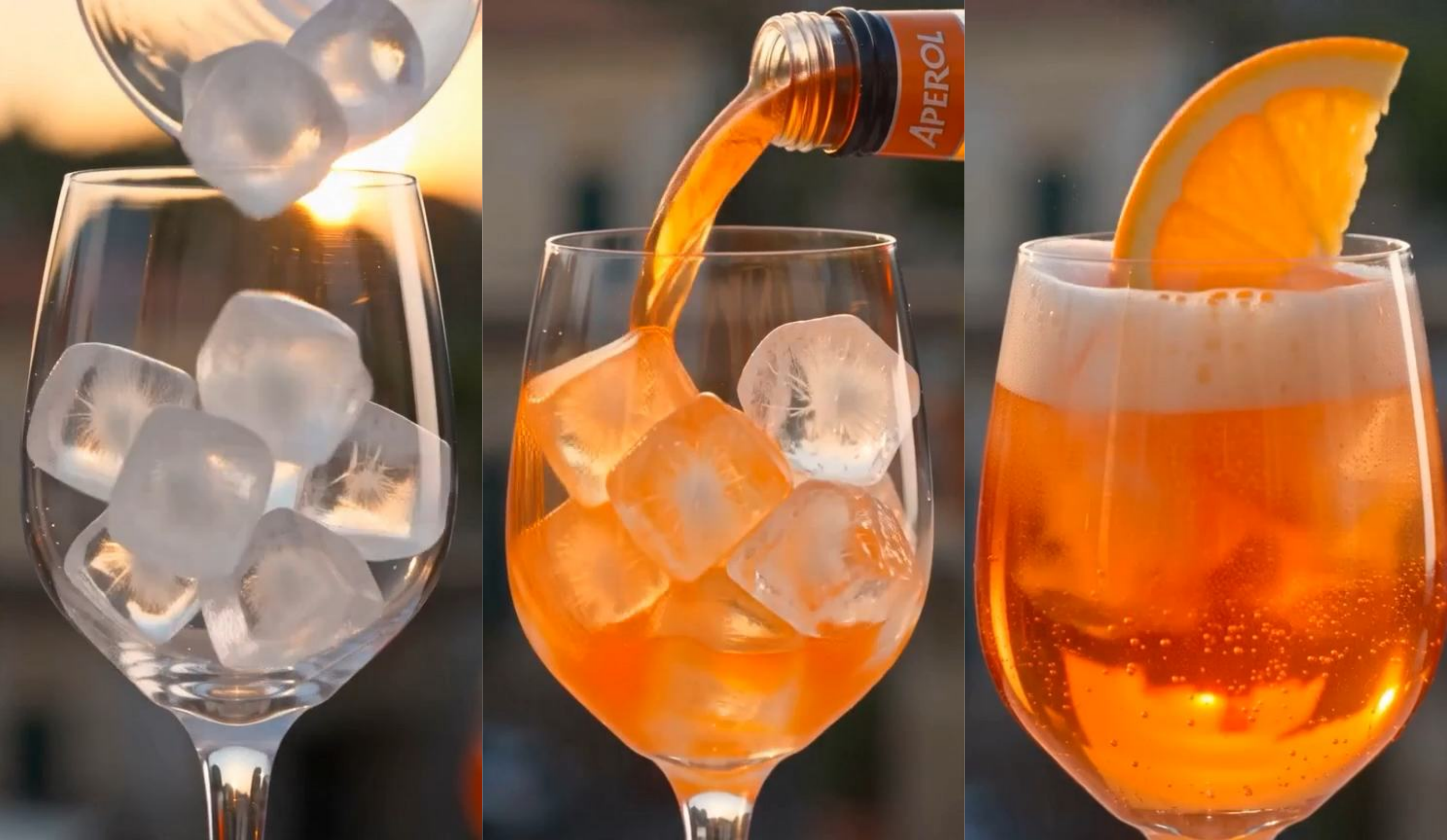}
  \caption{An Aperol-inspired ad showing ice cubes tumbling into a wine glass, Aperol cascading from a bottle, and an orange slice dropping in. The sequence feels one-dimensional.}
  \Description{A sequence of close-up product shots shows ice cubes falling into a glass, Aperol pouring from a bottle, and an orange slice dropping into the drink. All three frames use a similar centered composition and depict successive drink-preparation actions.}
  \label{fig:scene-aperol}
\end{figure}

An Aperol-inspired ad\footnote{Aperol-inspired cinematic ad: \url{https://youtu.be/eKJqJTd4myY}} features shots of a drink being prepared. It shows ice cubes tumbling into a wine glass, orange Aperol cascading from a bottle over ice, and an orange slice dropping in (Figure~\ref{fig:scene-aperol}). P5 thought these shots lacked diversity: \textit{``Shots from second 4 - 7 are all framed the exact same way. The ice in glass, then pouring of the Aperol, then the placement of the orange. There are 4 cuts in 3 seconds and each cut has the exact same framing. It makes this section one dimensional and uninteresting.''}

\subsubsection{Shot-to-Shot Continuity}

Does the connection between two adjacent shots feel continuous, coherent, and effective? All six participants discussed this dimension in 99 sentences across 39 videos. Participants commented on motion continuity, environmental continuity, and transition-device appropriateness.

\paragraph{Motion Continuity}
\label{sec:motion-continuity}

Do movement, action, and camera motion carry smoothly from one shot to the next? 

\begin{figure}
  \includegraphics[width=\columnwidth]{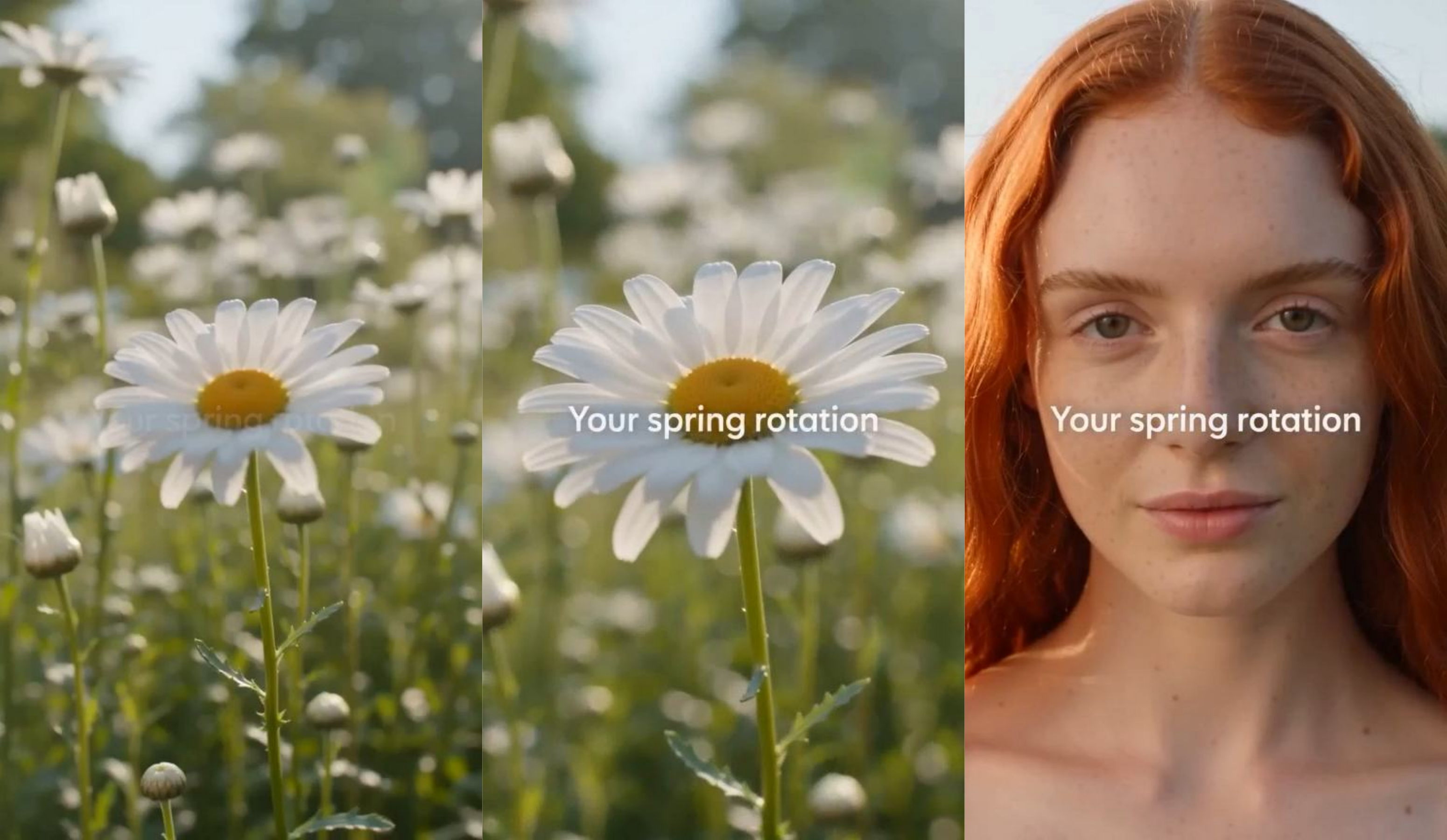}
  \caption{A Glossier-inspired ad. The first shot pushes in through a white chamomile flower (frames 1 and 2) while the second shot shows a model's face without camera movement (frame 3). P3 felt that the push-in should continue in the second shot for motion continuity.}
  \Description{Three frames from a Glossier-inspired ad. The first two move progressively closer to the center of a white chamomile flower. The third cuts to a static close-up of a model's face, interrupting the forward camera movement established by the flower shots.}
  \label{fig:motion-glossier}
\end{figure}

For instance, in a Glossier-inspired ad\footnote{Glossier-inspired cinematic ad: \url{https://youtu.be/cHLZik16Gcc}}, the first shot features a push-in through a white chamomile flower, which then dissolves into a second shot of a model's face (Figure~\ref{fig:motion-glossier}). However, the second shot remains entirely static. Highlighting this visual disconnect, P3 argued that the push-in \textit{``should have continued from the plant to the model's face''} to maintain a continuous camera motion across the transition.

\begin{figure}
  \includegraphics[width=\columnwidth]{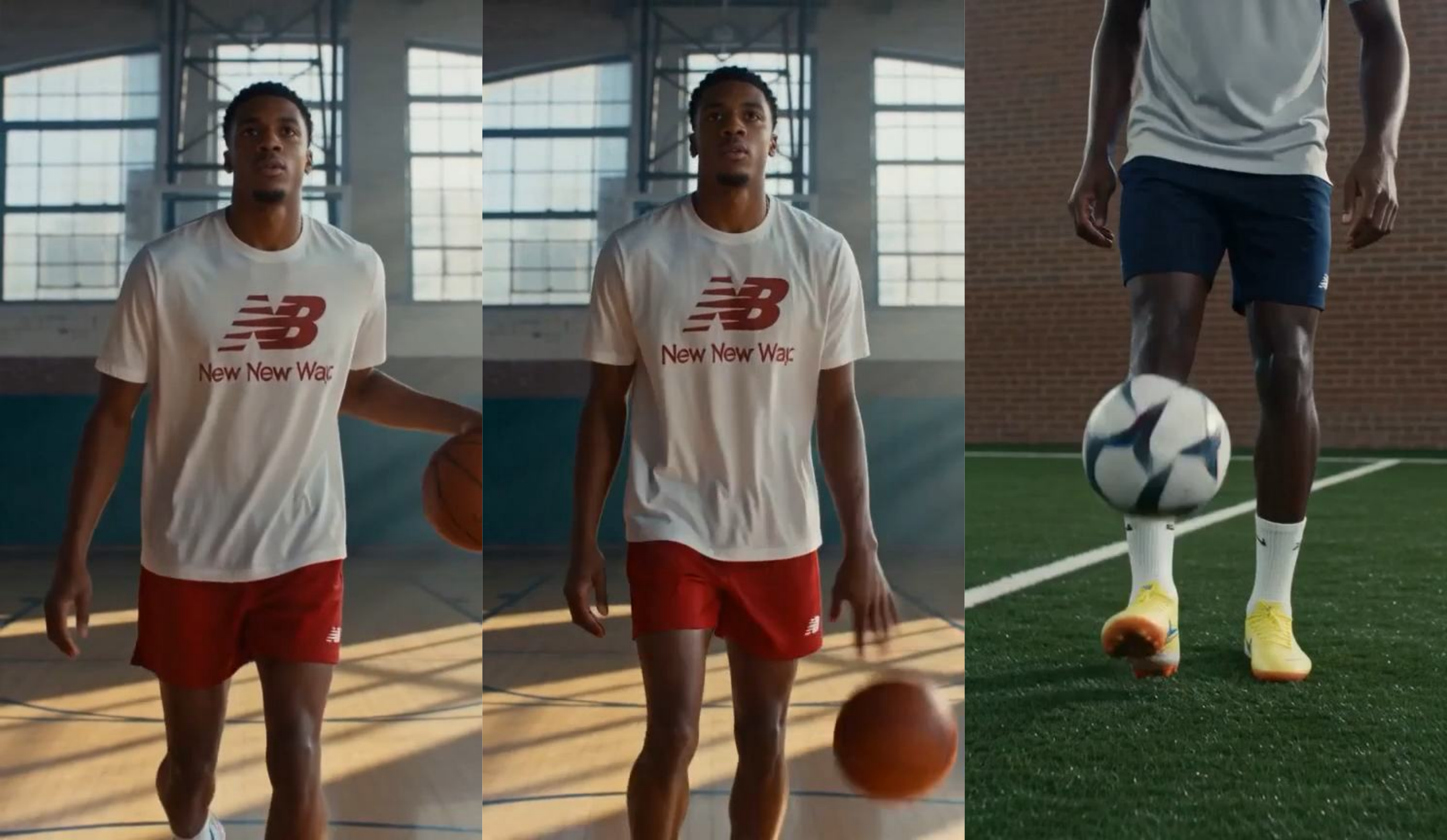}
  \caption{A New Balance-inspired ad showing a shot of a basketball player dribbling downward transitions into a shot of a soccer player flicking a soccer ball upward.}
  \Description{Frames from a New Balance-inspired ad show a basketball player pushing a basketball downward, followed by a soccer player flicking a soccer ball upward. The opposing ball movements align across the cut to form a movement match.}
  \label{fig:motion-new-balance}
\end{figure}

In a New Balance-inspired ad\footnote{New Balance-inspired cinematic ad: \url{https://youtu.be/DkaA8Ou4a0M}}, a shot of a basketball player dribbling downward transitions into a shot of a soccer player flicking a soccer ball upward (Figure~\ref{fig:motion-new-balance}). The transition is a movement match cut where the downward force of the basketball matches the upward lift of the soccer ball. P2 praised the execution and said, \textit{``I really like the basketball bounce sounds at 00:03 that suddenly change to soccer kicks [...] It really draws the viewer in and holds attention.''}

\begin{figure}
  \includegraphics[width=\columnwidth]{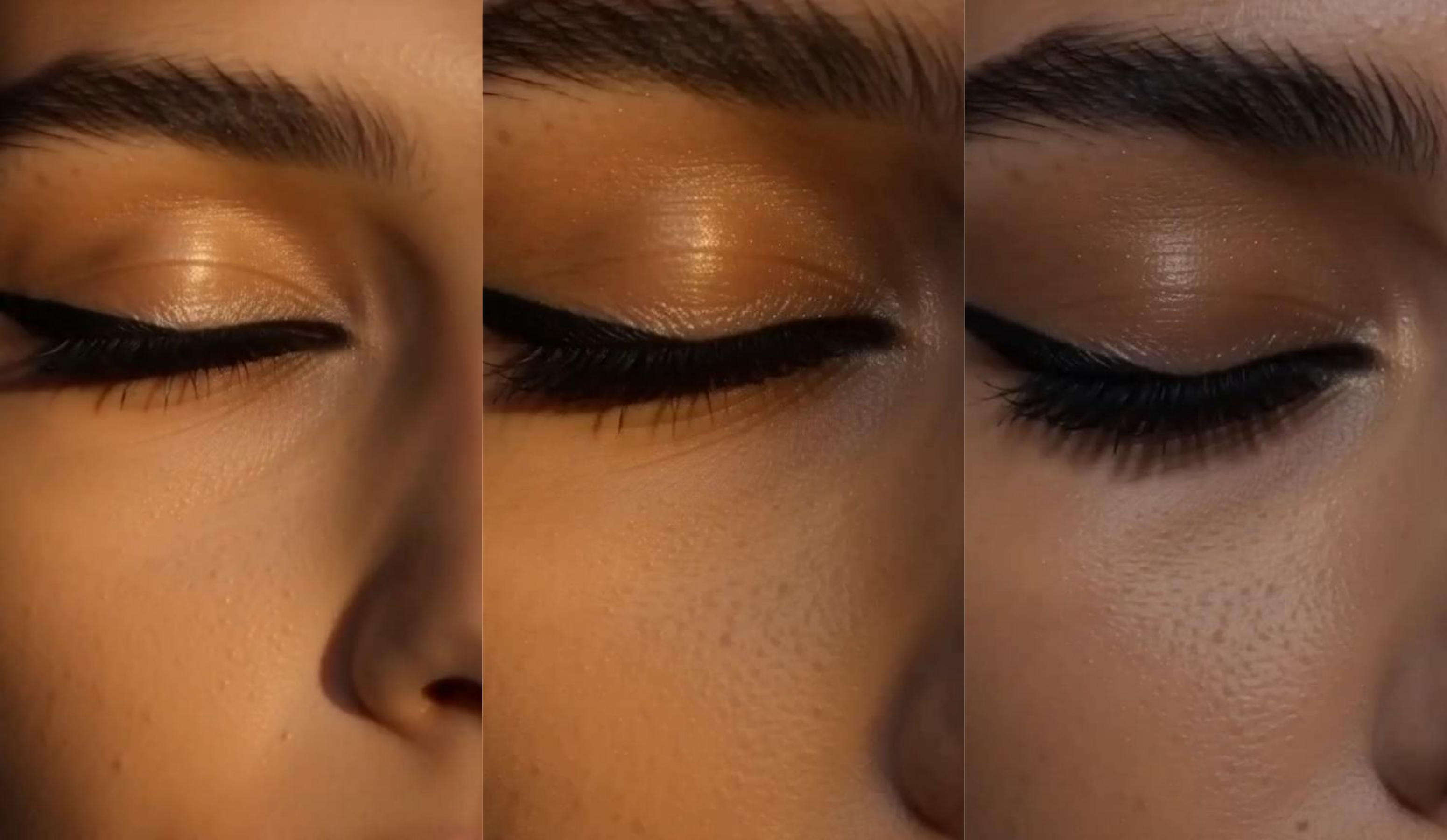}
  \caption{A Dior-inspired ad. It shows a sequence of three close-up shots of a model's eye. The camera position shifts only slightly between frames, resulting in a stuttering effect.}
  \Description{Three consecutive extreme close-ups of a model's eye. Each frame uses a slightly different camera position and crop while showing nearly the same subject, producing small discontinuous jumps across the sequence.}
  \label{fig:motion-dior}
\end{figure}

Four participants observed abrupt jump cuts within the ads. For example, a Dior-inspired ad\footnote{Dior-inspired cinematic ad: \url{https://youtu.be/DrbG_3l8uwA}} has a sequence of three shots showing the close-up of a model's eye (Figure~\ref{fig:motion-dior}). Across shots, the camera position shifts only slightly between frames, resulting in a stuttering effect. Reacting to this lack of visual continuity, P3 commented that the cuts feel \textit{``uncomfortable.''}

\paragraph{Environmental Continuity}

Do adjacent shots feel visually compatible in terms of setting, light, color, or scene context? 

\begin{figure}
  \includegraphics[width=\columnwidth]{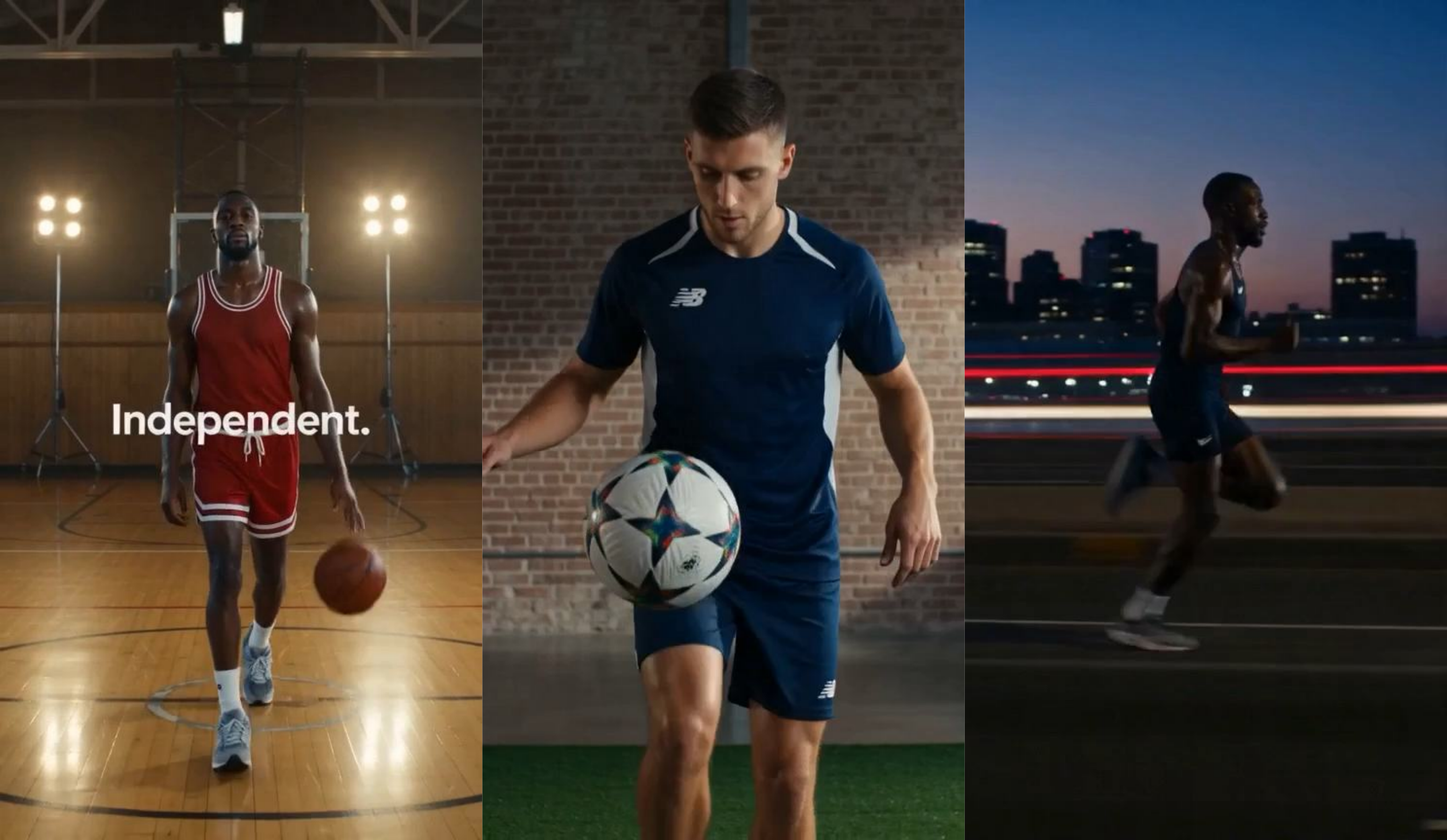}
  \caption{A New Balance-inspired ad jumping from a warm indoor gym to a bright soccer field and to a dark city skyline. The shift in environments feels jarring.}
  \Description{Three consecutive settings from a New Balance-inspired ad: a warmly lit indoor basketball gym, a bright green outdoor soccer field, and a dark blue city skyline at night. The lighting, colors, and environments change sharply between shots.}
  \label{fig:environmental-new-balance}
\end{figure}

In a New Balance-inspired ad\footnote{New Balance-inspired cinematic ad: \url{https://youtu.be/0JEvOoI70xU}}, P1 described the jump from a warm indoor gym to a bright soccer field and then to a dark city skyline as \textit{``jarring''} (Figure~\ref{fig:environmental-new-balance}). In their words, \textit{``The video jumps from a warm, nostalgic indoor gym to a bright green soccer field, and then suddenly to a dark, cold city skyline at night. Because the color grading and lighting setups are so radically different between these clips, the edit can feel a bit like a compilation of unrelated stock footage rather than a singular, cohesive commercial.''}

\paragraph{Transition-Device Appropriateness}

Does the transition (e.g., hard cut, dissolve, and wipe) between adjacent shots fit the tone of the ad?

\begin{figure}
  \includegraphics[width=\columnwidth]{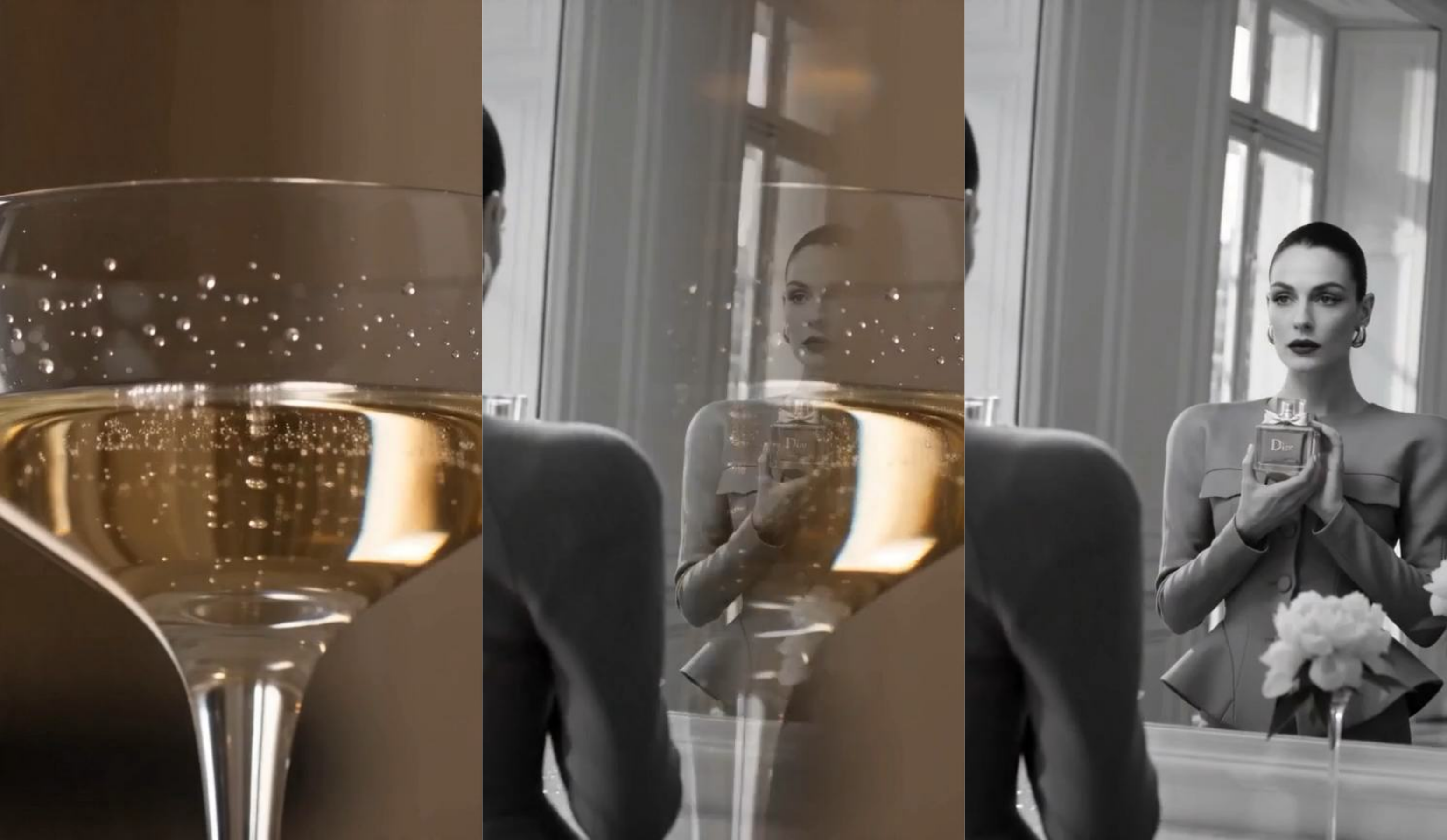}
  \caption{A Dior-inspired ad that uses a wipe transition. P4 commented that the wipe felt cheap.}
  \Description{Consecutive frames from a Dior-inspired ad illustrate a wipe transition, in which a hard moving boundary slides one image across the frame to replace the preceding image.}
  \label{fig:transition-dior}
\end{figure}

A Dior-inspired ad\footnote{Dior-inspired cinematic ad: \url{https://youtu.be/uFGQLCMDO50}} used a wipe transition between two shots (Figure~\ref{fig:transition-dior}). Feeling the wipe disrupted the brand's luxury aesthetic, P4 said, \textit{``Use of slide transition at 0:10 was jarring and felt like this was edited by a teen in iMovie.''}

Similarly, a Gucci-inspired ad\footnote{Gucci-inspired cinematic ad: \url{https://youtu.be/J6lTNlvM3rw}} uses a dissolve transition and P6 commented that they \textit{``don't like that dissolve effect''} and would \textit{``use a regular cut''} instead.

\subsubsection{Message and Brand Coherence}

Does the ad clearly communicate what is being advertised, what message viewers should take away, and whether the product, visuals, audio, and editing choices feel consistent with the intended brand? Five participants discussed this dimension in 97 sentences across 38 videos. Participants commented on message clarity, brand consistency, and product visibility.

\paragraph{Message Clarity}
\label{sec:message-clarity}

Can viewers infer what product is being advertised and what message the ad is trying to communicate, or does the sequence feel like a collection of disconnected shots?

\begin{figure}
  \includegraphics[width=\columnwidth]{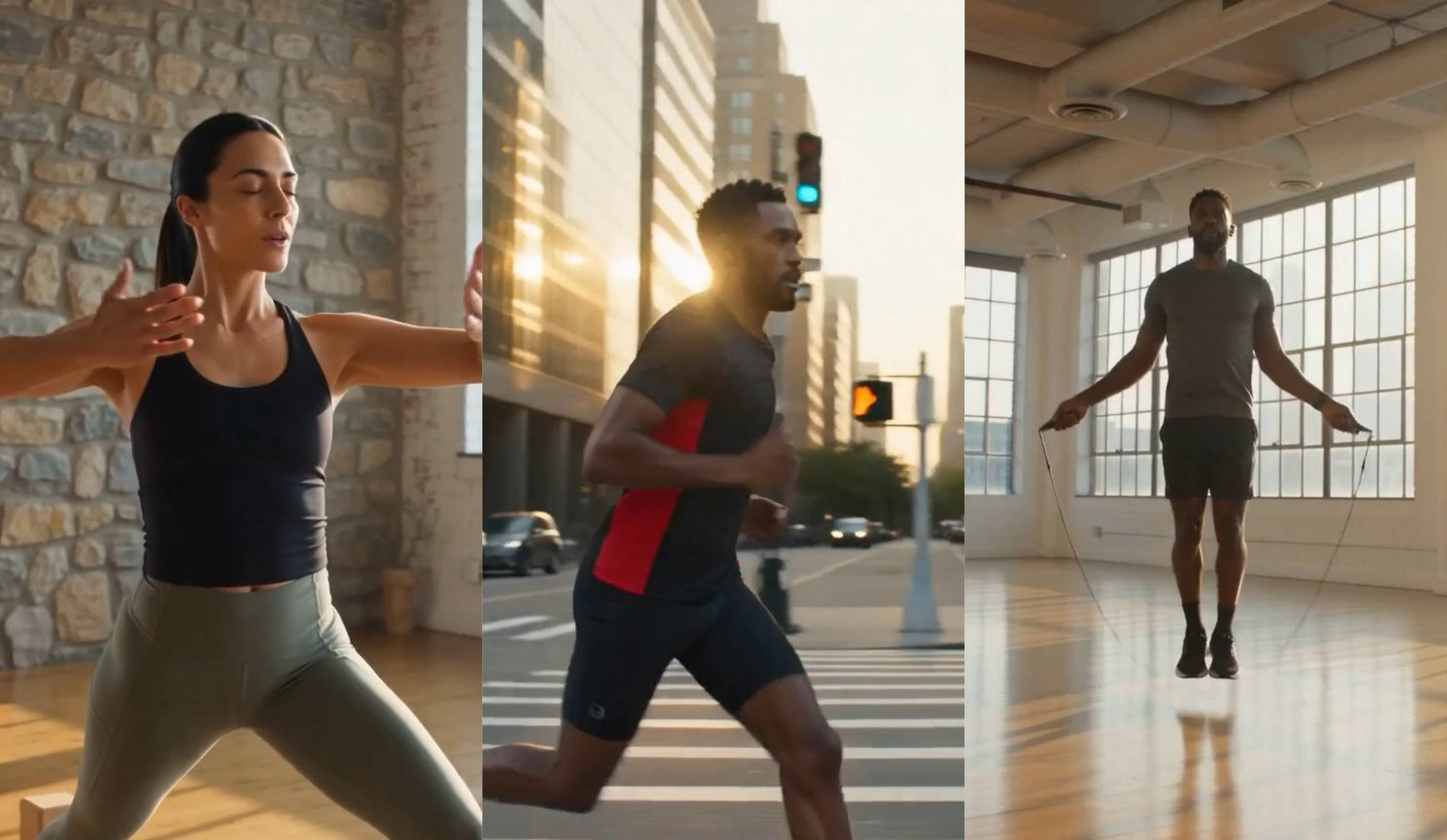}
  \caption{A Lululemon-inspired ad showing disconnected shots of a woman doing yoga, a runner sprinting, and a man jumping rope. The message being advertised is unclear.}
  \Description{Three disconnected activities in a Lululemon-inspired ad: a woman practicing yoga, a runner sprinting, and a man jumping rope. The frames do not show a shared product, setting, or clearly developing message.}
  \label{fig:mesage-lululemon}
\end{figure}

The use of montage can sometimes make the product or brand image being advertised confusing. For instance, a Lululemon-inspired ad\footnote{Lululemon-inspired cinematic ad: \url{https://youtu.be/2zRY6VZW0nc}} shows a montage of a woman doing yoga, a runner sprinting through a downtown intersection, and a man jumping rope (Figure~\ref{fig:mesage-lululemon}). P2 commented, \textit{``I have no clue what is being advertised here. There is relaxing yoga, then intense running, jumping rope, etc.''}

\begin{figure}
  \includegraphics[width=\columnwidth]{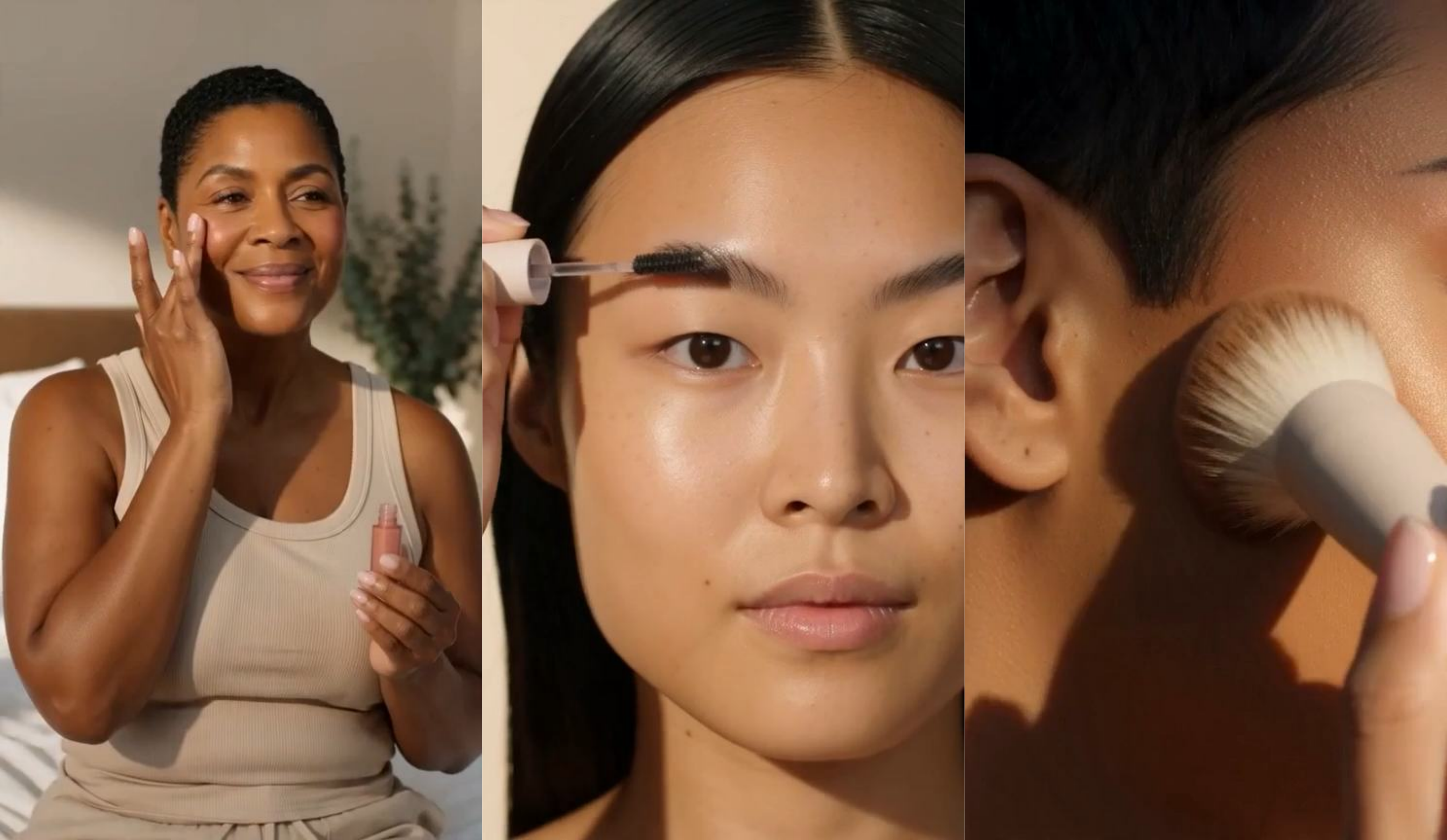}
  \caption{A Rare Beauty-inspired ad showing women using different products (liquid blush, clear brow gel, and bronzer brush), making it unclear what the ad is selling.}
  \Description{Frames from a Rare Beauty-inspired ad show different women using several products: liquid blush, clear brow gel, and a bronzer brush. No single product remains the consistent focus across the sequence.}
  \label{fig:message-rare-beauty}
\end{figure}

A Rare Beauty-inspired ad\footnote{Rare Beauty-inspired cinematic ad: \url{https://youtu.be/bvoscuYAbZ0}} shows a montage of a woman dabbing liquid blush onto her cheek with her fingertip, a woman applying clear brow gel with a spoolie, and a person sweeping a fluffy brush of bronzer along their cheekbone (Figure~\ref{fig:message-rare-beauty}). P4 found the sequence disorienting: \textit{``Overall, the edit is confusing, all different products are used, you don't know what this ad is for until the end. Again the tagline comes in at a very awkward point and is confusing what it is selling.''}

A Coca-Cola-inspired ad\footnote{Coca-Cola-inspired cinematic ad: \url{https://youtu.be/YghZTgPKecE}} shows a montage of two people clinking their glass Coca-Cola bottles, a skateboarder holding a bright red Coke while performing a kickflip, and a laughing guy tossing a plastic bottle of Coke. While P5 could tell that the message surrounded people having a good time drinking Coca-Cola, they felt the visuals failed to communicate a cohesive message: \textit{``I don't really understand what the narrative is here. I assume it's centered around people having a good time while drinking Coca Cola?''}

\paragraph{Brand Consistency}

Do the products, editing choices, visuals (e.g., logos and captions), and audio (e.g., tone) feel consistent with the intended brand rather than introducing conflicting or off-brand elements?

\begin{figure}
  \includegraphics[width=\columnwidth]{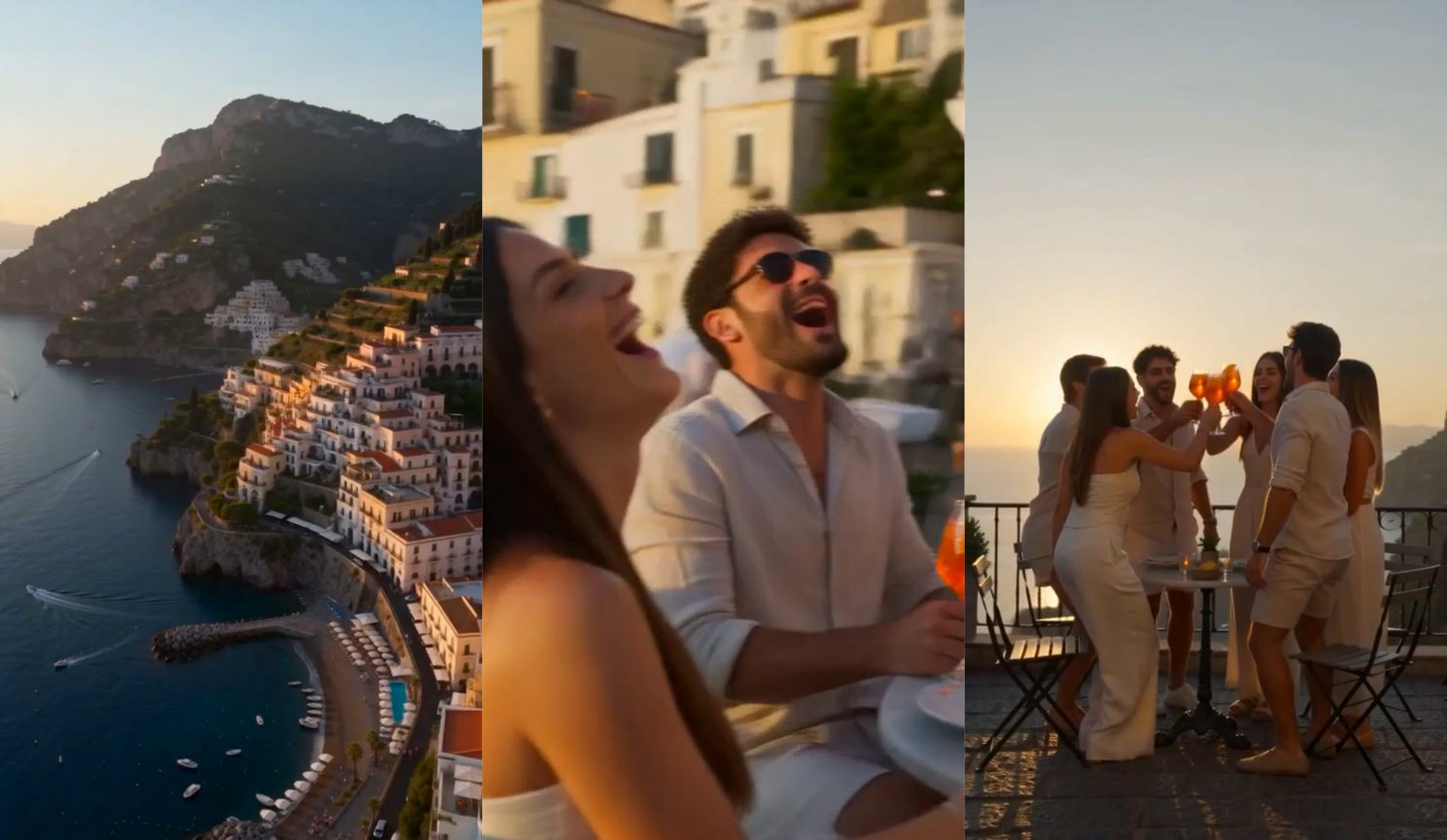}
  \caption{An Aperol-inspired ad showing a Mediterranean town and friends around a cafe table bursting into laughter. The scenes are consistent with Aperol's brand image.}
  \Description{Frames from an Aperol-inspired ad show a Mediterranean coastal town and a group of friends laughing together around a cafe table. Warm colors, outdoor socializing, and the coastal setting recur across the sequence.}
  \label{fig:brand-aperol}
\end{figure}

An Aperol-inspired ad\footnote{Aperol-inspired cinematic ad: \url{https://youtu.be/eKJqJTd4myY}} shows a Mediterranean town and friends around a cafe table bursting into laughter (Figure~\ref{fig:brand-aperol}). Both scenes are consistent with the Aperol brand. P5 said, \textit{``I like the opening shot of the Mediterranean village. I like the narrative direction of friends drinking and enjoying each other's company, and Aperol's relation to that.''}

\begin{figure}
  \includegraphics[width=\columnwidth]{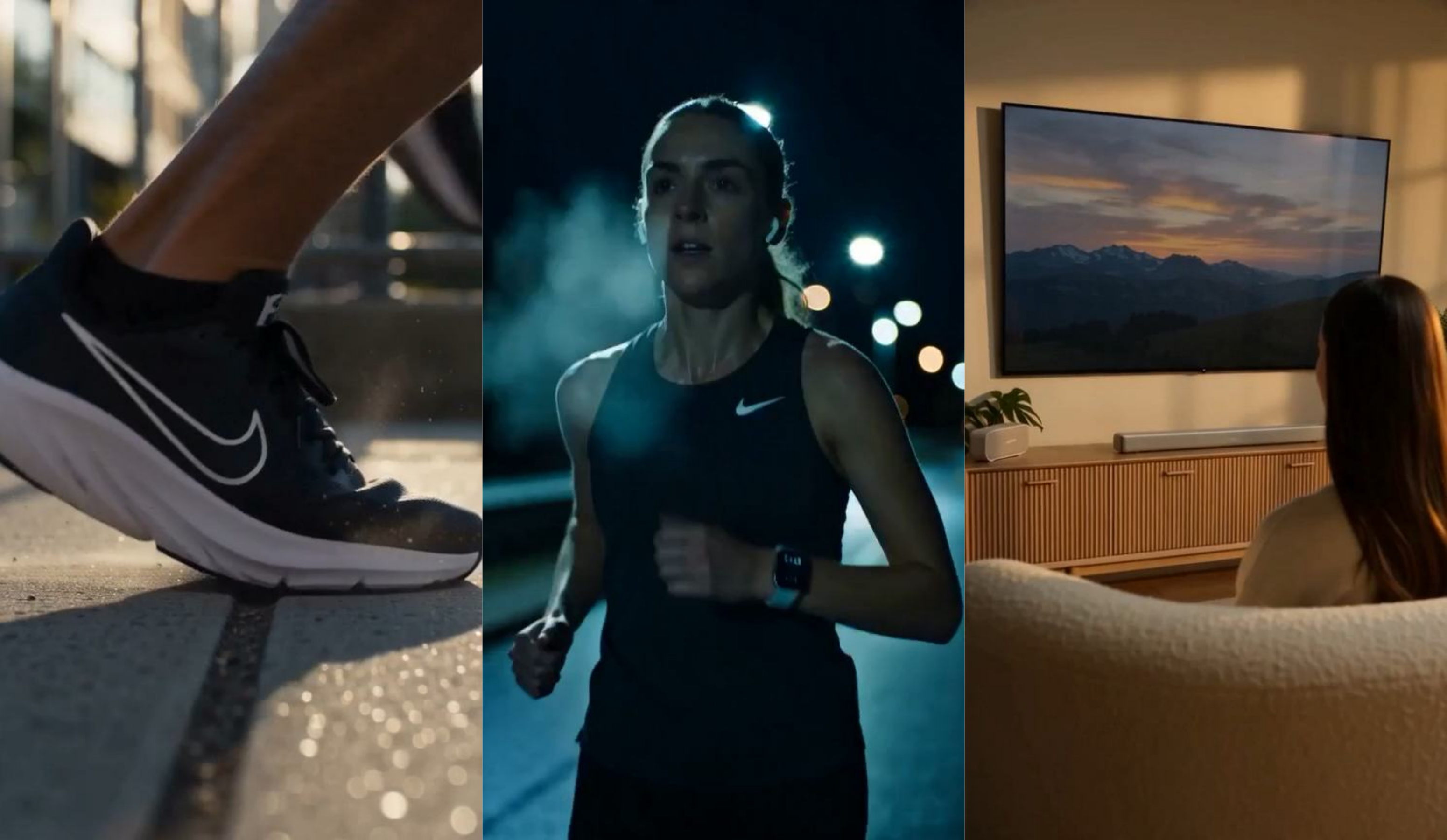}
  \caption{A Nike sneaker in a Lululemon-inspired ad (frame 1), A woman on a Nike shirt in an ad selling Apple products (frame 2), and a barely-visible Bose speaker at the bottom left of a television (frame 3).}
  \Description{Three examples of product and brand inconsistency. The first shows a Nike sneaker in a Lululemon-inspired ad. The second shows a woman wearing a Nike shirt in an ad for Apple products. The third shows a small Bose speaker partially obscured near the lower-left corner of a television.}
  \label{fig:brand-lululemon-apple-bose}
\end{figure}

In contrast, showing another brand’s logo creates a disconnect in an ad. For example, a Lululemon-inspired ad\footnote{Lululemon-inspired cinematic ad: \url{https://youtu.be/E0B7Kp-zt74}} shows a close-up of a Nike sneaker (Figure~\ref{fig:brand-lululemon-apple-bose} frame 1). P1 said, \textit{``The video opens on the woman practicing yoga on a rug, but at 0:02, the editor cuts to a close-up of a foot stepping down in a white-swoosh Nike sneaker, immediately followed by the woman back in her yoga flow [...] Mixing a competitors' easily identifiable shoe into an edit that ends on a Lululemon logo creates a confusing brand message.''}

Similarly, an Apple-inspired ad\footnote{Apple-inspired cinematic ad: \url{https://youtu.be/3VqC9pb3YMs}} shows a person running with an Apple Watch. Her shirt has a Nike logo (Figure~\ref{fig:brand-lululemon-apple-bose} frame 2). P2 said, \textit{``I would definitely remove the Nike logo from her shirt at 00:09. It doesn't belong in a commercial about Apple.''}

\paragraph{Product Visibility}

Is the product or key product details (e.g., logo) visible enough at the moments when the viewer needs to recognize it?

For example, in a Bose-inspired ad\footnote{Bose-inspired cinematic ad: \url{https://youtu.be/i8VS8vZbJPs}}, a Bose speaker sits at the bottom left of a television (Figure~\ref{fig:brand-lululemon-apple-bose} frame 3). The product looks small and the logo is illegible. P2 commented that \textit{``in the couch behind-the-scenes shot, the product isn't visible enough.''}

\subsubsection{Temporal Rhythm and Pacing}

Do the pacing and speed changes create an effective rhythm for the ad’s intended energy and emotional effect? All six participants discussed this dimension in 64 sentences across 30 videos. Participants commented on pacing and speed manipulation.

\paragraph{Pacing}

Does the pacing feel right for the energy and emotional response the ad intends to communicate?

All participants commented on the pacing of the ads: \textit{``the pacing is strong in this!''} (P4), \textit{``the pacing of the video is on point''} (P2), and \textit{``the pacing is so far the best along with Apple's video''} (P3).

Fewer shots could communicate a sense of relaxation. An Alo Yoga-inspired ad\footnote{Alo Yoga-inspired cinematic ad: \url{https://youtu.be/WImCoQ9OH_s}} shows a montage of a woman doing yoga. It has few cuts to create a relaxing atmosphere. P2 said, \textit{``I think the pacing works well for this one. You don't want to have too many shots in something that is yoga related since it is supposed to be a relaxing activity.''}

Rapid shots could build a high energy. The Under Armour-inspired ad\footnote{Under Armour-inspired cinematic ad: \url{https://youtu.be/rmEU0QCFUh8}} (Figure~\ref{fig:plan}) uses rapid montage to communicate the feeling of high-intensity workouts. P1 noted, \textit{``The ad utilizes a quick succession of shots showing different aspects of grind and athleticism (stair running, shoe tying, lifting, catching). This rapid pacing builds momentum and mimics a high-intensity workout.''}

Rapid cuts could disrupt the relaxing tone of an ad. A Casper-inspired ad\footnote{Casper-inspired cinematic ad: \url{https://youtu.be/vHHMBhhKrng}} conveys sleep as a relaxing activity, but there are too many cuts: \textit{``The jump cuts are too quick - you want this ad to feel relaxing, but the jump cuts make it too intense''} (P4).

\paragraph{Speed Manipulation}

Does the use of time-based effects (e.g., slow motion, speed ramps, and fast motion) feel intentional and effective?

\begin{figure}
  \includegraphics[width=\columnwidth]{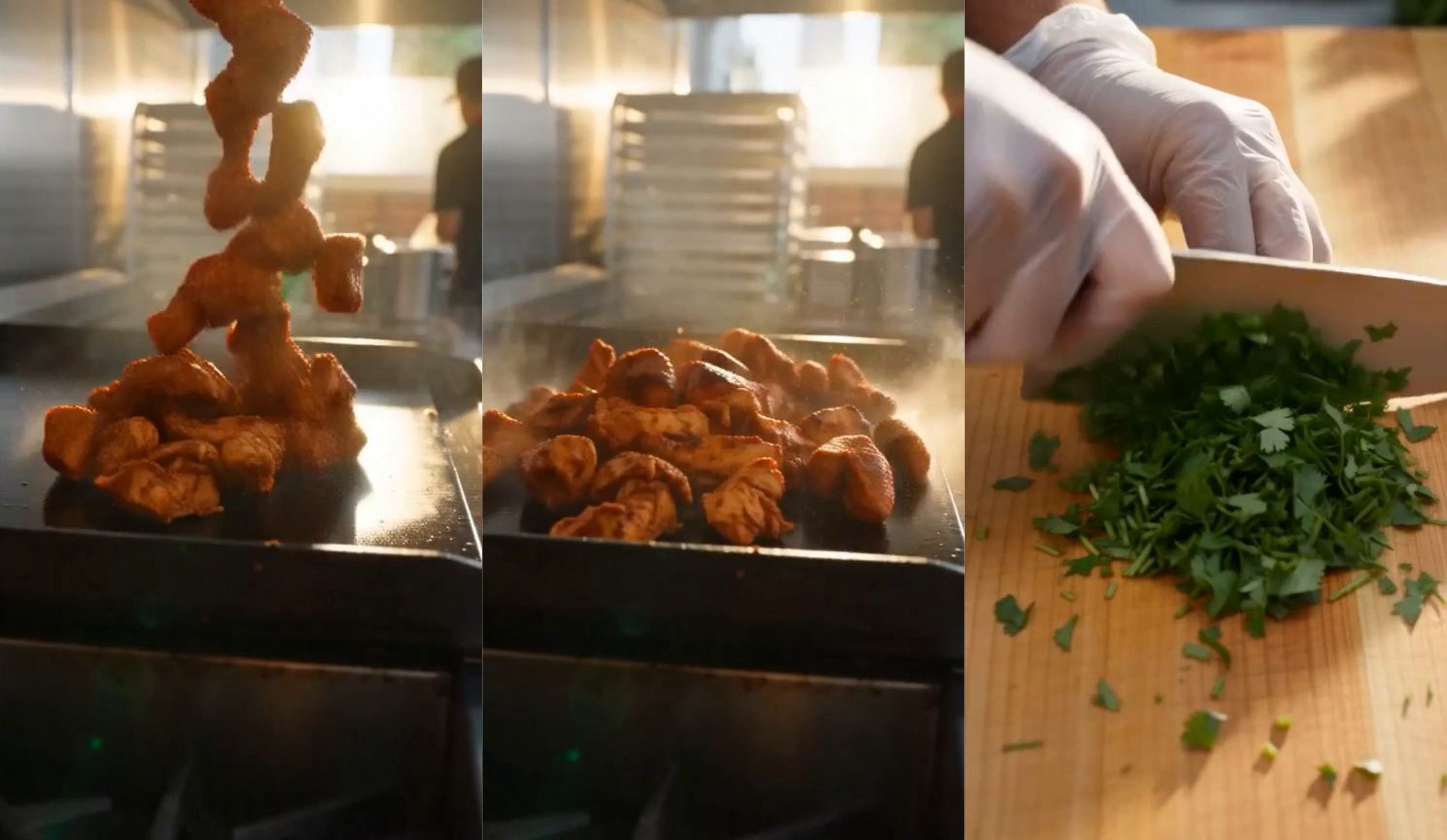}
  \caption{A Chipotle-inspired ad showing marinated chicken landing on a hot plancha and chopping fresh cilantro on a wooden board. The first shot uses a fast-to-slow speed ramp, but P6 suggested reversing it to slow-to-fast.}
  \Description{Frames from a Chipotle-inspired food-preparation sequence show marinated chicken landing on a hot plancha and fresh cilantro being chopped on a wooden board. The chicken action is presented as successive stages of a fast-to-slow speed ramp.}
  \label{fig:speed-chipotle}
\end{figure}

In a Chipotle-inspired ad\footnote{Chipotle-inspired cinematic ad: \url{https://youtu.be/WcdElS9w1hs}}, the first shot shows marinated chicken landing on a hot plancha and the second shot shows chopping fresh cilantro on a wooden board (Figure~\ref{fig:speed-chipotle}). The first shot utilized a fast-to-slow speed ramp. However, P6 suggested reversing the effect to a slow-to-fast speed ramp so the accelerated landing of the chicken in the first shot seamlessly bridges into the downward chopping motion in the second: \textit{``I'd start with a slow to fast motion speed ramp on the first clip and make a movement match cut at the end of that clip.''}

\subsubsection{Generation Defects and Others}

Besides critiquing the editing quality, participants also commented on generation artifacts in Q4 (additional comments about this advertisement beyond its editing). We coded these sentences as generation defects. For example, in a Mercedes-Benz-inspired ad\footnote{Mercedes-Benz-inspired cinematic ad: \url{https://youtu.be/MdbasrT_fXg}}, P2 noted that \textit{``the car is driving backwards.''} In a Ralph Lauren-inspired ad\footnote{Ralph Lauren-inspired cinematic ad: \url{https://youtu.be/kVR50sBaRBo}}, P6 observed that \textit{``the lady's head is backwards while she is walking forwards.''}

We coded sentences that did not fit the six dimensions and generation defects as others. For example, P6 was not able to pinpoint any weaknesses in a Ralph Lauren-inspired ad and answered \textit{``no comments''} for Q2 (editing choices that feel weak, ineffective, or missing).

\section{Discussion}

Here, we discuss how our dimensions suggest directions for improving, evaluating, and automatically assessing AI-generated cinematic ads. Finally, we discuss the connections between our framework and the literature and the limitations of our work.

\subsection{Improving AI-Generated Cinematic Ads}

In Study 2, professional editors critiqued the editing choices that felt weak and ineffective. These failure modes highlight directions for future research. For example, many editing problems involve shot-to-shot motion discontinuity (Section~\ref{sec:motion-continuity}): while individual shots look polished, adjacent shots fail to connect through motion, action, and camera movement. Another example is a lack of clarity in the intended message (Section~\ref{sec:message-clarity}): while the adjacent shots might connect well, the overall collection of shots and montage might fail to communicate the product being advertised and the message viewers should take away.

One potential way to improve AI-generated cinematic ads is editing-aware planning. Instead of generating a video directly from a prompt, future video generation systems could first generate a shot plan, critique the plan along the six dimensions, and revise the plan before rendering. For example, a planner could emphasize that consecutive shots should preserve motion continuity in the plan, decide on a core product message to convey, and evaluate whether the planned montage supports the message.

\subsection{Towards Human Evaluation of AI-Generated Cinematic Ads}

The six dimensions can also support more systematic human evaluation of AI-generated cinematic ads. During the human evaluation of these ads, raters are provided with some ads and score each ad using a questionnaire.

One immediate use of our dimensions is rubric design. The evaluation questionnaire will contain rubrics for the six dimensions and the sub-dimensions. With the rubrics, rather than ask raters for a single overall editing-quality score, future studies could ask raters to provide more fine-grained judgment. Such rubrics would make human evaluation more diagnostic: two videos could receive similar overall scores but fail for different reasons.

Our findings in Study 2 can also be used as training materials for human raters. For future research, we consolidated the findings from Study 2 in a document in the supplementary materials. This document contains the definitions of the six dimensions, the definitions of the observed sub-categories, and up to two bad examples and two good examples for each sub-category.

\subsection{Towards Automated Evaluation of AI-Generated Cinematic Ads}

During the automated evaluation of cinematic ads’ editing quality, an autorater, such as a multimodal large language model (MLLM), scores ads using some rubrics. It is critical to ensure that the autoraters’ judgments align with those of human raters. To do so, researchers often collect a prompt set and generate a video for each prompt. Both the autorater and human raters then score the ads. Researchers then calculate the correlation between the two sets of scores to assess the alignment~\cite{song-etal-2025-vf}. 

Beyond rubric design, our framework provides the operational definitions for the six dimensions and the sub-categories when prompting the autorater. For example, such definitions could be incorporated into the autorater’s system prompt. Encoding the definitions explicitly may help the autorater produce ratings that are better aligned with human raters trained using the same framework.

\subsection{Connections to Existing Literature}

Some of our dimensions have roots in film theories and media aesthetics. This convergence suggests that our participants drew on established craft knowledge rather than idiosyncratic preferences. Shot-to-shot continuity is central to Murch's ``Rule of Six''~\cite{murch2001blink} and Smith's attentional theory of cinematic continuity~\cite{smith2012attentional}; temporal rhythm and pacing recalls Pearlman's account of editing rhythm~\cite{pearlman2016cutting}; audiovisual coordination reflects Chion's analysis of synchronization and silence~\cite{chion1994audiovision}; visual composition connects to Zettl's applied media aesthetics~\cite{zettl2017sight}. Our contribution lies in showing that these principles transfer to a new evaluation target of cinematic ads.

Focusing on cinematic ads, our framework also extends beyond existing film theories. First, message and brand coherence has little counterpart in film literature and is absent from film-oriented frameworks such as FilmEval~\cite{2025arXiv250618899H}. Instead, it aligns more closely with marketing theories. For example, Keller's brand equity framework emphasizes a consistent brand image across touchpoints~\cite{keller1993conceptualizing}. Second, our participants often evaluated narrative progression as the coherence of a montage rather than adherence to a dramatic arc~\cite{d5b5dba5-a060-358d-bcca-a85d7586b5b3}. This could be specific to cinematic ads since they do not have a problem–resolution narrative structure. 

\subsection{Limitations}

Our findings should be interpreted in light of the study scope. The generated ads were produced using a specific pipeline: LLM-based shot planning followed by video rendering with a selected video-generation model. Other pipelines may produce different strengths and weaknesses. For example, a model that generates a video directly from a single prompt may fail in different ways than a model that follows a structured shot plan. However, the dimensions we identified could serve as an initial framework for evaluating editing quality.

\section{Conclusion}

This paper examined how professional video editors evaluate the editing quality of AI-generated cinematic ads. In Study 1a, we characterized ad formats on social media and established cinematic ads as a common format. In Study 1b, we further analyzed the duration, shot count, audio and text elements, and editing techniques in cinematic ads. We observed that cinematic ads on social media tended to be shorter than typical television ads and to have shorter shot durations than movies. Grounded in Studies 1a and 1b, we created a two-step pipeline for generating cinematic ads. We then generated 70 AI-produced cinematic ads, asked professional editors to critique what worked, what failed, and how they would revise the ads, and derived six dimensions of editing quality from the critiques. The resulting six-dimension framework offers a vocabulary for diagnosing editing problems in AI-generated cinematic ads and provides a foundation for future systems that plan, evaluate, and revise generated videos with greater awareness of professional editing practice. This work also contributes to research in evaluating AI-generated videos by showcasing how to derive evaluation criteria using a human-centered approach.

\section*{Third-Party Content and Trademark Notice}

The third-party advertisements referenced in this paper remain the copyrighted property of their respective rights holders. Brand names, product names, trademarks, logos, and selected advertisement frames are used solely for identification, scholarly analysis, and criticism. The authors claim no ownership of these third-party materials and are not affiliated with, sponsored by, or endorsed by the brands discussed in this paper.

The AI-generated videos produced for Study 2 are research stimuli and are not official advertisements. They were created solely for noncommercial research and evaluation. Their inclusion of, or reference to, a brand name or logo does not imply affiliation with or endorsement by the corresponding brand.

\bibliographystyle{ACM-Reference-Format}
\bibliography{references}

\end{document}